\documentclass[a4paper]{ar-1col}

\jname{Ann. Rev. Nuc. Part. Sc.}
\jvol{AA}
\jyear{2016}

\usepackage{graphicx}
\usepackage{bm}
\usepackage{color}
\usepackage{amsmath}
\usepackage{amssymb}
\usepackage{cancel}
\usepackage[numbers,sort&compress]{natbib}

\usepackage[linktocpage=true]{hyperref}
\hypersetup{
colorlinks=true,
citecolor=fignumcolor,
linkcolor=headcolor,
urlcolor=fignumcolor,
pdfauthor={},
pdftitle={},
pdfsubject={}
}

\newcommand{\be}{\begin{equation}}
\newcommand{\ee}{\end{equation}}
\newcommand{\bea}{\begin{eqnarray}}
\newcommand{\eea}{\end{eqnarray}}
\def\ba#1\ea{\begin{align}#1\end{align}}
\newcommand{\refeq}[1]{Eq.~(\ref{eq:#1})}          
\newcommand{\refeqs}[2]{Eqs.~(\ref{eq:#1})--(\ref{eq:#2})}          
\newcommand{\reffig}[1]{Fig.~\ref{fig:#1}}          

\newcommand{\refsec}[1]{Sec.~\ref{sec:#1}}

\newcommand{\vs}{\nonumber\\}

\newcommand{\vk}{\bm{k}}
\newcommand{\vx}{\bm{x}}

\newcommand{\rhob}{\overline{\rho}_m}
\newcommand{\rhoDE}{\overline{\rho}_{\rm DE}}
\newcommand{\Om}{\Omega_m}

\renewcommand{\d}{\delta}

\renewcommand{\O}{\mathcal{O}}

\renewcommand{\v}[1]{\bm{#1}}

\newcommand{\iMpch}{\,h\,{\rm Mpc}^{-1}}

\newcommand{\vn}{\bm{\nabla}}

\def\Mpl{M_{\rm Pl}} 
\def\PhiN{\Phi_{\rm N}} 
\def\Phim{\Phi_{\gamma}} 
\def\gsim{\, \lower .75ex \hbox{$\sim$} \llap{\raise .27ex \hbox{$>$}} \,}
\def\lsim{\, \lower .75ex \hbox{$\sim$} \llap{\raise .27ex \hbox{$<$}} \,}

\newcommand{\mentry}[1]{{\color{fignumcolor} {\sffamily\bfseries#1}}}

\DeclareFontFamily{OT1}{pzc}{}
\DeclareFontShape{OT1}{pzc}{m}{it}%
             {<-> s * [1.00] pzcmi7t}{}
\DeclareMathAlphabet{\mathscr}{OT1}{pzc}%
                                 {m}{it}

\begin{document}

\markboth{Joyce, Lombriser, Schmidt}{Dark Energy vs. Modified Gravity}

\title{Dark Energy vs. Modified~Gravity}

\author{Austin Joyce,$^1$ Lucas Lombriser,$^2$ and Fabian~Schmidt$^3$\vspace{.2cm}
\affil{$^1$Enrico Fermi Institute and Kavli Institute for Cosmological Physics,
University of Chicago, Chicago, IL 60637; email: ajoy@uchicago.edu}\vspace{.2cm}
\affil{$^2$Institute for Astronomy, University of Edinburgh, Royal Observatory, Blackford Hill, Edinburgh, EH9 3HJ, U.K.; email: llo@roe.ac.uk}\vspace{.2cm}
\affil{$^3$Max-Planck-Institute for Astrophysics, D-85748 Garching, Germany; email: fabians@mpa-garching.mpg.de}}


\begin{abstract}
Understanding the reason for the observed accelerated expansion of the Universe represents one of the fundamental open questions in physics.  In cosmology, a classification has emerged among physical models for the acceleration, distinguishing between \emph{Dark Energy} and \emph{Modified Gravity}.  In this review, we give a brief overview of models in both categories as well as their phenomenology and characteristic observable signatures in cosmology.  We also introduce a rigorous distinction between Dark Energy and Modified Gravity based on the strong and weak equivalence principles.
\end{abstract}

\maketitle

\tableofcontents

\section{INTRODUCTION}
\label{sec:intro}

The accelerated expansion of the Universe, discovered in 1998 \cite{Riess:1998cb,Perlmutter:1998np}, has raised fascinating questions for cosmology and physics
as a whole.  Elucidating the physics behind the acceleration is one of the main science drivers for a major experimental effort using large-scale structure (LSS) surveys.  
All current observations are consistent with a cosmological constant (CC); while this is in some sense the most economical possibility, the CC has its own theoretical and naturalness problems~\cite{Weinberg:1988cp,Martin:2012bt}, so it is worthwhile to consider alternatives.
A convenient classification scheme has emerged
in the field of cosmology, separating different physical scenarios
of accelerating cosmologies into the categories of ``Dark Energy''
and ``Modified Gravity''.  Essentially, Dark Energy models
modify the stress-energy content of the Universe, adding an additional 
component with equation of state $w \simeq -1$.  That is, we modify
the right-hand side of the Einstein equations.  The Modified Gravity
category corresponds to modifying the left-hand side, {\it i.e.}, General Relativity (GR)
itself, for example by modifying the Einstein--Hilbert action.  
This review intends to provide an overview of these two categories, and describe
the cosmological observables and tests that are able to distinguish
experimentally between these two scenarios of physics driving
the acceleration.

As we will see, while there are models which unambiguously belong to one category or the other,
in reality there is a continuum of models between the two extremes
of ``pure'' Dark Energy and Modified Gravity such that a
strict division into these two categories is to some extent a matter of
personal preference.  In this review, we will base the distinction
in theory space upon the Strong Equivalence Principle (SEP).  A compatible, slightly
less rigorous, but more
practical observational distinction in cosmology is to call Modified Gravity
those models which feature a light universally coupled degree of freedom
mediating a fifth force.

Throughout we restrict ourselves to observations on 
extragalactic scales, which translates to scales of order
1 Mpc and above. Further, we only consider scenarios for which effects become
relevant at late times.  
That is, we disregard the possible presence of such effects in the 
early Universe (often referred to as ``early Dark Energy'').  

We begin in \refsec{theory} with an overview of the 
landscape of Dark Energy and Modified Gravity models.  \refsec{pheno} then describes the cosmological
phenomelogy of these models.  \refsec{tests} provides a summary of the 
observables and experimental methods used in cosmology to test Dark Energy and Modified Gravity.  We conclude with an
outlook in \refsec{outlook}.  Throughout, we work in units where $c = \hbar = 1$.

\section{OVERVIEW: DARK ENERGY AND MODIFIED GRAVITY}
\label{sec:theory}

In this section, we briefly review Dark Energy (DE) and Modified
Gravity (MG) models before turning to the tricky issue of drawing a boundary between
these two paradigms. We will not attempt to be comprehensive---there exist many excellent reviews on both subjects, for example~\cite{Frieman:2008sn,Clifton:2011jh}---but rather want to give a flavor of the types of models and their phenomenology.

It is worth mentioning at the outset that the models we describe are not {\it per se} solutions of the CC problem. 
Generally, it is still necessary to invoke some other physics or symmetry principle to explain why the expected contribution to the CC from Standard Model fields is absent. 
Despite lacking a complete solution of the CC problem, the general approaches considered here are promising and represent reasonable possibilities for how physics in the gravitational sector may behave.

\subsection{Dark Energy (DE)}
\label{sec:quintessence}

Perhaps the most natural direction to explore to explain the observed value of the CC is to posit that the CC is itself a dynamical field, dubbed Dark Energy, which relaxes to its present value through some mechanism~\cite{Wetterich:1987fm,Peebles:1987ek,Ratra:1987rm}. 
There are many possible incarnations of this idea, what they share is the presence of a degree of freedom which develops a condensate profile that drives the background cosmological evolution.  
The simplest one is to imagine that the CC is a canonical scalar field with a potential\footnote{We define the reduced Planck mass as $M_{\rm Pl}^2  = (8\pi G)^{-1}$.}
\be
S = \int{\rm d}^4x\sqrt{-g}\left(\frac{M_{\rm Pl}^2R}{2} -\frac{1}{2}(\partial\phi)^2-V(\phi)\right).
\label{eq:canonicalDE}
\ee
We will refer to this model as {\it quintessence}~\cite{Caldwell:1997ii}.\footnote{In his influential review~\cite{Weinberg:1988cp}, Weinberg showed that models of quintessence necessarily have to be tuned in order to make the CC small.}  
A homogeneous condensate of this field, $\phi = \phi(t)$, will behave as a perfect fluid, with equation of state, $w = P/\rho$, given by
\be
w = \frac{\frac{1}{2}\dot\phi^2-V(\phi)}{\frac{1}{2}\dot\phi^2+V(\phi)}\,,
\ee
where here and throughout, dots denote derivatives with respect to time.  
Observations restrict $w$ to be very close to $-1$ in the present universe (note that this need not be the case for all times). This is equivalent to the requirement that the evolution of the field be potential-dominated. In this sense, the present stage of cosmic acceleration in quintessence models is similar to a period of very low scale inflation. It is also important to note that the quintessence equation of state is generally time-dependent.
\begin{marginnote}
The \mentry{phenomenology of quintessence models} is rich; however, their qualitative behavior can be separated into two types~\cite{Caldwell:2005tm}: models that are just starting to deviate from $w \simeq -1$ (thawing), and models which are approaching $w\simeq-1$ today (freezing).
\end{marginnote}

An intriguing aspect of quintessence models is that a wide class of models exhibit ``tracking" behavior~\cite{Wetterich:1994bg,Copeland:1997et,Ferreira:1997au,Zlatev:1998tr}, where the energy density in the field closely traces the energy density in dark matter until the near past, providing a possible explanation of the coincidence problem---that is, why the current density of DE and dark matter are comparable. 
 The canonical example is the Ratra--Peebles potential~\cite{Ratra:1987rm}
\be
V(\phi) = \frac{M^{4+n}}{\phi^n},
\ee
with $n > 0$.

Another consideration to keep in mind with DE (and even to some extent MG) models is to be sure that we are not secretly reintroducing a CC. 
As a simple example, a scalar field with an exactly flat potential could plausibly be thought of as DE, but in this case we should really absorb this constant potential into the bare CC in the lagrangian. Another compelling possibility to consider is that DE and dark matter may be ``unified" into a single component. However, this turns out to be so fine-tuned that one effectively goes back to a two-fluid model~\cite{Sandvik:2002jz}.

\subsubsection{More general DE models}
Even within the paradigm of the CC as a dynamical scalar field, there is a plethora of different models, with varied phenomenology. 
Perhaps the most obvious generalization away from~\refeq{canonicalDE} is to consider a non-canonical scalar field---one which possesses derivative self-interactions. 
Concretely, these models may be written as
\be
S = \int{\rm d}^4x\sqrt{-g} \Lambda^4 K(X)~,
\label{eq:kxlag}
\ee
where we have defined $X\equiv -\frac{1}{2\Lambda^4}(\partial\phi)^2$ and $K(X)$ is an arbitrary function.
Models of this type are often referred to as $k$-essence~\cite{Chiba:1999ka,ArmendarizPicon:2000ah}. 
Similar to the canonical case, for homogeneous field profiles, $\phi = \phi(t)$, these models behave as a perfect fluid, though of a more general type, where now the equation of state parameter $w$ is given by
\be
w = \frac{K}{2XK_{,X}-K}~.
\ee
By suitably choosing a functional form for $K$, it is possible to reproduce the cosmic expansion history. Field perturbations away from the background evolution no longer propagate luminally in $k$-essence models, but rather propagate with a speed of sound given by
\be
c_s^2 = \frac{K_{,X}}{K_{,X}+2XK_{,XX}},
\ee
this allows structure formation to be noticeably different with respect to canonical DE models, as we discuss in~\refsec{bsmDE}.

DE models of the late-time acceleration are similar to inflation, and there has been an effort to adapt the effective field theory of inflation formalism~\cite{Cheung:2007st} to the problems of DE and MG~\cite{Creminelli:2008wc,Gubitosi:2012hu,Bloomfield:2012ff,Gleyzes:2013ooa,bellini/sawicki}.

\begin{textbox}
\section{Dark Energy vs. Modified Gravity:} In the text, we draw a distinction between Dark Energy and Modified Gravity by means of the {\color{headcolor} {\sffamily\bfseries strong equivalence principle (SEP)}}. We classify any theory which obeys the SEP as Dark Energy, and any theory which violates it as Modified Gravity (see \refsec{dab}). 
Heuristically, the strong equivalence principle forbids the presence of fifth forces.
\end{textbox}

\subsection{Modified Gravity (MG)} \label{sec:MG}
We begin by reviewing some commonly-studied infrared modifications of GR.  These examples will allow us to draw some more general lessons: generic modifications of gravity lead to new physics on small scales~\cite{Lue:2003ky}. Indeed, all known modifications to Einstein gravity introduce new degrees of freedom in the gravitational sector.\footnote{In this review, we focus on modified gravity applications to \emph{cosmic acceleration}. There has also been work to modify gravity to obviate the need for dark matter, but we will not discuss this.}  
These new particles mediate a fifth force, so the theory must employ some {\it screening mechanism} in order to evade local tests of gravity, which are very constraining.

\subsubsection{Scalar-tensor theories}
Scalar-tensor theories are probably the best-studied modifications to Einstein gravity. Restricting to canonical scalar fields for the moment, these theories can be cast in Einstein frame as
\be
S = \int{\rm d}^4x\sqrt{-g}\left(\frac{M^2_{\rm Pl}R}{2}-\frac{1}{2}(\partial\phi)^2-V(\phi)\right)+S_{\rm matter}\left[A^2(\phi)g_{\mu\nu}, \psi\right].
\label{eq:STaction}
\ee
Here $R$ is the Ricci scalar constructed from the metric $g_{\mu\nu}$. Matter fields, $\psi$, couple to the Jordan-frame metric $\tilde g_{\mu\nu} = A^2(\phi)g_{\mu\nu}$. This coupling induces interactions between matter fields and $\phi$, causing test particles in the Newtonian regime to feel a force:
\be
{\bm a} =  -\vn\bigg(\Phi + \ln A(\phi)\bigg),
\label{eq:acc}
\ee
sourced by both the Einstein-frame potential, $\Phi$, and the scalar, $\phi$. 

The prototypical scalar-tensor theory is Brans--Dicke theory~\cite{Brans:1961sx}, corresponding to $V(\phi) = 0$ and $A^2(\phi) = \exp[-\phi/(M_{\rm Pl}\sqrt{3/2+\omega})]$ in~\refeq{STaction}; the theory is most often cast in Jordan frame, $g_{\mu\nu} \mapsto \tilde g_{\mu\nu}$, where it takes the form 
\be
S = \frac{M_{\rm Pl}^2}{2}\int{\rm d}^4x\sqrt{-\tilde g}\left(\phi \tilde R-\frac{\omega}{\phi} (\partial\phi)^2\right)+S_{\rm matter}\left[\tilde g_{\mu\nu},\psi\right],
\label{eq:JframeBD}
\ee
with $\omega$ a constant parameter. Solar System tests of gravity place the bound $\omega~ \gsim~4\times 10^{4}$ \cite{Clifton:2011jh,Will:2014kxa}. This corresponds to taking the coupling to matter to be very weak, making the Brans--Dicke theory essentially equivalent to a DE model.

In more general scalar-tensor theories---by suitably choosing $V(\phi)$ and $A(\phi)$---it is possible to evade Solar System constraints, while still having interesting phenomenology for the scalar. 
To see how this can work, we focus on the usual case where the scalar field $\phi$ couples to the trace of the Jordan-frame stress tensor, $\tilde T = A^{-4} T$, where $T$ is the trace of the stress tensor in Einstein frame. 
If we consider non-relativistic sources, $T = -\rho$, and define $\bar\rho \equiv A^{-1}\rho$, the scalar obeys the equation of motion
\be
\square\phi = \frac{{\rm d}V_{\rm eff}}{{\rm d}\phi}~~~~~~~~~~~~~~V_{\rm eff}(\phi) = V(\phi)+A(\phi)\bar\rho~.
\ee

The key point is that $\phi$ responds to an {\it effective} potential, which depends on the external matter sources. 
This makes it possible to engineer situations where the field behaves differently in different environments. 

The most well-known example is the chameleon field~\cite{Khoury:2003aq,Khoury:2003rn}, where the potential and matter coupling are chosen so that the effective mass of the scalar field:
\be
m^2_{\rm eff}(\phi) = \frac{{\rm d}^2V_{\rm eff}}{{\rm d}\phi^2} = \frac{{\rm d}^2V}{{\rm d}\phi^2}+ \frac{{\rm d}^2A}{{\rm d}\phi^2}\bar \rho,
\ee
increases in regions of high density, like the Solar System. 
A concrete example is the Ratra--Peebles potential and a linear coupling~\cite{Khoury:2003aq,Khoury:2003rn}
\be
V(\phi) = \frac{M^{4+n}}{\phi^n}~~~~~~~~~~A(\phi) \simeq 1 + \xi \frac{\phi}{M_{\rm Pl}}.
\ee

Another popular scalar-tensor model with a screening mechanism is the symmetron model~\cite{Hinterbichler:2010es}, which can also be cast as~\refeq{STaction}, but with
\be
V(\phi) = -\frac{\mu^2}{2}\phi^2+\frac{\lambda}{4}\phi^4~~~~~~~~~A(\phi) \simeq 1 + \frac{\phi^2}{2M^2}.
\label{eq:Vsymmetron}
\ee
In regions of low density, the ${\mathbb Z}_2$ symmetry of the model is spontaneously broken and $\phi$ mediates a fifth force; in regions of high density, the symmetry is restored and the additional force turns off.  
More precisely, in both the chameleon and symmetron a screened field is sourced only by a {\it thin shell} near the surface of a dense object.  Whether or not the screening operates depends on the Newtonian potential of the object exceeding a certain threshold~\cite{Khoury:2003rn,Hinterbichler:2010es} (see \refsec{screening}). 

One of the most-studied scalar-tensor theories is so-called $f(R)$ gravity, which on the face of it is not a scalar-tensor theory at all. This model consists of higher-curvature corrections to the Einstein--Hilbert action
\be
S = \frac{M^2_{\rm Pl}}{2}\int{\rm d}^4x\sqrt{-g}\Big(R+f(R)\Big) + S_{\rm matter}[g_{\mu\nu}, \psi],
\label{fofraction}
\ee
where $f(R)$ is a function only of the Ricci scalar, chosen to become significant in the low-curvature regime $R\to 0$.  
In~\cite{Capozziello:2002rd,Capozziello:2003tk,Carroll:2003wy}, $f(R)$ models were used to drive cosmic acceleration. 
However, it was shown in~\cite{Erickcek:2006vf}, that these original models are actually in conflict with precision tests of gravity. 
This is essentially because these models are scalar-tensor theories in disguise~\cite{Barrow:1988xh,Chiba:2003ir}, which can be seen by performing a field redefinition and conformal transformation (see {\it e.g.},~\cite{Joyce:2014kja} for details), to cast the theory as in~\refeq{STaction} with
\be
V(\phi) = \frac{M_{\rm Pl}}{2}\frac{\left(\phi f_{,\phi}-f\right)}{\left(1+f_{,\phi}\right)^2}~~~~~~~~~A(\phi)  = e^{\frac{\phi}{\sqrt{6}M_{\rm Pl}}},
\ee
which is equivalent to the Brans--Dicke theory~\refeq{JframeBD} with $\omega = 0$ and a potential.

In~\cite{Hu:2007nk,Starobinsky:2007hu}, $f(R)$ models compatible with local tests of gravity were constructed, for example in~\cite{Hu:2007nk} $f(R)$ is given by 
\be
f(R) = -m^2\frac{c_1\left(\frac{R}{m^2}\right)^n}{c_2\left(\frac{R}{m^2}\right)^n+1} \approx \rho_{\Lambda,\rm eff} + f_{R0} \left(\frac{R}{\bar R_0}\right)^{-n}\,,
\label{eq:husawickifr}
\ee
where the second form is the leading expression for $R/m^2 \ll 1$ which is required by cosmological and local tests (\refsec{tests}).  The parameters $c_1,\,c_2$ can be adjusted so that $\rho_{\Lambda,eff}$ supplies a CC consistent with observations, leaving $f_{R0}$ as a free parameter.  $\bar R_0$ is the background Ricci scalar today, so that $f_{R0}$ corresponds to the field value in the background today. 
In the scalar-tensor language, the model is consistent with experiment because of the chameleon mechanism. $f(R)$ gravity has been explored in a variety of contexts, and we refer the reader to~\cite{DeFelice:2010aj} for more of the specifics.

\subsubsection{More general scalar-tensor theories: Horndeski and generalizations}
The space of scalar-tensor theories is far broader than the example we have presented in~\refeq{STaction}. In fact, in recent years, there has been an effort to map out the most general consistent theory of a metric interacting with a scalar. In this context, consistency requires that the theory be absent of ghosts. Typically, theories which have equations of motion which are higher than second order in time derivatives have a ghost.\footnote{This is the content of Ostrogradsky's theorem~\cite{Woodard:2015zca}.} Therefore, much of the interest has focused on theories which have second-order equations of motion.

In~\cite{Deffayet:2011gz} the most general theory of a scalar field interacting with a metric which has second-order equations of motion was derived, corresponding to the lagrangian
\begin{marginnote}
A \mentry{ghost} is a quantum with either negative energy density or negative norm. The existence of propagating ghosts in a theory poses problems for its quantization. In particular, the vacuum is unstable to decay into ghosts and healthy particles~\cite{Carroll:2003st,Cline:2003gs}.
\end{marginnote}
\begin{align}
\nonumber
{\cal L}_{\rm gen. gal.} =&~ K(\phi, X)-G_3(\phi, X)\square\phi+G_4(\phi, X)R\\
&+G_{4, X}(\phi, X)\big[(\square\phi)^2-(\nabla_\mu\nabla_\nu\phi)^2\Big]
+G_5(\phi, X)G_{\mu\nu}\nabla^\mu\nabla^\nu\phi\label{eq:Horndeski}\\\nonumber
&-\frac{1}{6}G_{5, X}(\phi, X)\Big[(\square\phi)^3-3 (\square\phi)(\nabla_\mu\nabla_\nu)^2+2\nabla^\mu\nabla_\alpha\phi\nabla^\alpha\nabla_\beta\phi\nabla^\beta\nabla_\mu\phi\Big].
\end{align}
This theory has four arbitrary functions, $K, G_3, G_4, G_5$; and $X$ is defined as in~\refeq{kxlag}. Interestingly, an equivalent theory had been derived much earlier by Horndeski~\cite{Horndeski:1974wa,Kobayashi:2011nu}, but had gone essentially unnoticed in the literature. 

A particularly well-studied limit of this theory is the so-called galileon~\cite{Nicolis:2008in}; the simplest version of this theory---the cubic galileon---is a special case of \refeq{Horndeski},
\be
{\cal L}_{\rm gal.} = \frac{M_{\rm Pl}^2R}{2}-\frac{1}{2}(\partial\phi)^2 - \frac{1}{\Lambda^3}\square\phi(\partial\phi)^2.
\label{eq:cubicG}
\ee
In the limit where gravity is taken to be non-dynamical, this theory has the symmetry $\delta\phi = c + b_\mu x^\mu$. In general dimensions, there are only a finite number of terms invariant under this symmetry, which also have second-order equations of motion. 
The galileon appears in many constructions; in particular it describes the decoupling limits of both massive gravity~\cite{deRham:2010ik} and of the Dvali--Gabadadze--Porrati (DGP) model~\cite{Dvali:2000hr,Luty:2003vm}. For a review of many interesting properties of the galileon, see~\cite{deRham:2012az}.

Recently, an effort has been made to relax the assumption of second-order equations of motion in order to construct the most general scalar-tensor type theory which propagates three degrees of freedom nonlinearly. In these theories, the expected ghostly degrees of freedom from higher-order equations of motion are projected out by constraints (see~\cite{Langlois:2015cwa} and references therein).

\subsubsection{Massive gravity and braneworlds}
\label{sec:massive}
Einstein gravity is the theory of a {\it massless} spin-2 particle. A natural question to ask is whether it is possible for the graviton to be a massive particle. The intuition behind this explanation for cosmic acceleration is that massive fields induce potentials of the Yukawa-type
\be
V(r) \sim \frac{e^{-m r}}{r},
\ee
which shut off at distances of order $m^{-1}$. The idea is that the presently-observed cosmic acceleration could be due to a weakening of gravity at large scales, with $m \sim H_0$.

Massive gravity has been studied since the initial investigations of Fierz and Pauli~\cite{Fierz:1939ix}, but much of the modern interest is due to the construction by de Rham-Gabadadze-Tolley of a ghost-free theory \cite{deRham:2010ik,deRham:2010kj} (building on~\cite{ArkaniHamed:2002sp,Creminelli:2005qk}), 
which showed how to overcome the traditional difficulties associated with nonlinearly completing the Fierz--Pauli theory. In particular, generic nonlinear theories propagate a ghost in addition to the graviton degrees of freedom~\cite{Boulware:1973my}. The absence of this Boulware--Deser ghost in the dRGT theory was shown in~\cite{Hassan:2011hr}. Recovery of GR in the massless limit relies on the Vainshtein mechanism (\refsec{screening}). For reviews of the theoretical background and recent developments, see~\cite{Hinterbichler:2011tt,deRham:2014zqa}.

Another avenue to massive gravity is as a resonance (as opposed to the hard mass above). This idea has been very influential in cosmology; the most studied example being the DGP model~\cite{Dvali:2000hr}. The set-up consists of a 4-dimensional brane (on which matter fields live) embedded in an infinitely large 5-dimensional bulk spacetime. The model consists of dynamical gravity both on the brane and in the bulk:
\be
S = \frac{M_5^3}{2}\int{\rm d}^5X\sqrt{-G}{\cal R}+\frac{M_4^3}{2}\int{\rm d}^4x\sqrt{-g}R,
\ee
with $M_5, M_4$ the 5d and 4d Planck masses, $G,{\cal R}$ the 5d metric and Ricci scalar, and $g, R$ the 4d metric and Ricci scalar. 
From the perspective of an observer living on the brane, gravity is mediated by a continuum of gravitons, and at short distances, gravity appears 4-dimensional, but at large distances, the gravitational potential ``leaks" off the brane and is that of a 5-dimensional theory. The potential due to a mass $M$ source behaves as~\cite{Deffayet:2001pu,Hinterbichler:2011tt}
\be
V(r)\simeq \left\{
\arraycolsep=1.4pt\def\arraystretch{1.9}
\begin{array}{l}
{\displaystyle\frac{M}{M_4^2 r}~~~~~~~~~~~~~~~r \ll r_c}\\\vspace{.02cm}
{\displaystyle\frac{M}{M_4^2 r}\frac{r_c}{r}~~~~~~~~~~~~r \gg r_c}
\end{array}
\right.,
\ee
where $r_c = M_4^2/2 M_5^3$.
We therefore see that the potential is {\it weaker} at large distances than it would be in pure GR.  
This weakening allows for de Sitter solutions on the brane absent of a bare CC~\cite{Deffayet:2000uy,Deffayet:2001pu}.
Unfortunately, this self-accelerating branch of solutions is unstable---perturbations are ghostly~\cite{Luty:2003vm,Nicolis:2004qq}.
Nevertheless, the DGP model remains quite interesting as a benchmark model; it is one of the better-studied MG models in the literature. For a review of DGP, see~\cite{Lue:2005ya}.

Interestingly, a massive graviton could address the CC problem by means of {\it degravitation}~\cite{Dvali:2002pe,ArkaniHamed:2002fu}: a large CC does not gravitate because a massive graviton behaves as a high-pass filter~\cite{Dvali:2007kt}.  However, a realistic degravitating solution has not yet been found in either dRGT or braneworld.

\subsubsection{Screening mechanisms}
\label{sec:screening}
In all of the models we have discussed, new degrees of freedom appear in the gravitational sector. 
This is in fact a general feature: GR is the unique low-energy theory of a massless spin-2 particle~\cite{Weinberg:1965rz,Feynman:1996kb}, so essentially any departure from GR introduces new degrees of freedom, 
typically with a mass of order Hubble today $m\sim H_0$.
Since they couple to the Standard Model, these light fields mediate a long range force between matter sources. 

The presence of additional forces is strongly constrained by laboratory and Solar System tests~\cite{Will:2014kxa}. 
Therefore, the additional degrees of freedom must hide themselves locally. 
The {\it screening mechanisms} by which this is achieved fall into several broad classes, activating in regions where the Newtonian potential $\PhiN$ or successive gradients become large.  $\PhiN$ is defined as the potential which solves the Poisson equation,~\refeq{Poisson}.
For concreteness, we assume that the additional degree of freedom is a scalar in the following discussion, but the general philosophy is much broader. For a more complete discussion, see~\cite{Joyce:2014kja}.  

These classes of models are described as follows.
\begin{itemize}
\item
{\color{fignumcolor}\bf Screening at large field values:}  The first type of screening we will consider activates in regions where the Newtonian potential exceeds some threshold value, $\PhiN > \Lambda$. 
Generally this leads to the additional degree of freedom itself developing a large vacuum expectation value---which causes the coupling to matter to weaken, the mass of the field to increase, or the self-coupling of the field to become large---leading to a diminishing of the force mediated. 
Notable examples are the chameleon~\cite{Khoury:2003aq,Khoury:2003rn}, symmetron~\cite{Hinterbichler:2010es,Pietroni:2005pv,Olive:2007aj} and dilaton~\cite{Damour:1994zq,Brax:2010gi}. 
From a phenomenological viewpoint, these mechanisms activate in regions of large potential, $\PhiN$, so regions of small Newtonian potential should exhibit the largest deviations from GR. 
\vspace{.1cm}
\begin{marginnote}
On subhorizon scales in cosmology, and within the Compton wavelength of
the field, $\phi$, and the Newtonian potential, $\PhiN$, are
proportional in the absence of screening, as may be deduced using the
scalar EOM and the Poisson equation.
\end{marginnote}
\item {\color{fignumcolor}\bf Screening with first derivatives:} This mechanism turns on when the local gravitational acceleration exceeds some critical value: $\vn\PhiN > \Lambda^2$. 
This condition roughly corresponds to first gradients of the scalar field becoming large; screening occurs due to kinetic self-interactions: $\partial\phi/\Lambda^2 \gg 1$. 
General $P(X)$ models can display this kind of screening~\cite{deRham:2014wfa}, with a broad class going by the name $k$-mouflage~\cite{Babichev:2009ee}.  
Since these models screen in regions of large acceleration, it is intriguing to imagine that they could be relevant for reproducing the phenomenological successes of MOND~\cite{Milgrom:1983ca} in a more complete framework~\cite{Babichev:2011kq,Khoury:2014tka}.
\vspace{.1cm}
\item {\color{fignumcolor}\bf Screening with second derivatives:} The last category of screening mechanisms we discuss is those which become active in regions of large curvature: $\nabla^2\PhiN > \Lambda^3$, which is equivalent to high density. 
These mechanisms rely on nonlinearities in the second derivatives: $\square\phi/\Lambda^3 \gg 1$. 
The most commonly-studied example in this class is the Vainshtein mechanism~\cite{Vainshtein:1972sx}, which operates in the galileon and in massive gravity.
In these models, the largest deviations from GR are expected in low curvature regimes. 
Additionally, there is evidence that Vainshtein screening is less efficient than naive estimates in time-dependent scenarios~\cite{deRham:2012fw,Chu:2012kz}, making this a promising avenue to search for deviations from GR.
\end{itemize}

\subsection{Drawing a Boundary}
\label{sec:dab}
In order to meaningfully discuss observationally discriminating between MG and DE, we must draw a distinction between the two scenarios. 
This is a somewhat aesthetic choice, but nonetheless helps us to organize our thinking about exploring and testing the various possibilities.

The distinction we make relies essentially upon the motion of bodies in the theory. To begin with, we recall the {\it weak equivalence principle} (WEP). 
The WEP is the statement that there exists some (usually Jordan-frame) metric to which all matter species couple universally. Then, test bodies---regardless of their composition---fall along geodesics of this metric.
This is usually stated as the equivalence of inertial and gravitational mass.
In this review, we focus on theories which satisfy the WEP \emph{at the level of the action}.

\begin{marginnote}
\mentry{Test bodies} are objects which are sufficiently small that we may neglect gravitational tidal forces in deriving their motion. A more precise definition can be found in~\cite{gralla:2010}.
\end{marginnote}

To distinguish DE {\it vs.}~MG, we further invoke the {\it strong equivalence principle} (SEP). 
The SEP extends the universality of free fall to massive bodies, {\it i.e.} to be completely independent of a body's composition, including gravitational binding energy, so compact objects like black holes also follow geodesics~\cite{1993tegp.book.....W}.
We call anything which obeys the SEP DE, and anything which does not, MG.
The motivation for this definition is to classify models which influence ordinary matter only gravitationally as DE. 
In these models, the force felt between two bodies is only that of GR (and possibly other Standard Model forces). 
However, in models of MG, bodies may carry additional charges ({\it e.g.}, scalar charge) which leads to them experiencing an additional force beyond that of gravity.
The appeal to the SEP is an attempt to make this intuition precise.  
A theoretical motivation for this distinction based upon the SEP is that it is believed---though not proven---that GR is the only metric theory which obeys the SEP~\cite{1993tegp.book.....W,Will:2014kxa}. 

The preceding discussion is somewhat abstract, so it is useful to illustrate the main points with scalar-tensor theory. We again consider the action~\refeq{STaction}, but allow for each matter species to have a different coupling to $\phi$,
\be
S = \int{\rm d}^4x\sqrt{-g}\left(\frac{M^2_{\rm Pl}}{2}R-\frac{1}{2}(\partial\phi)^2-V(\phi)\right)+S_{\rm matter}\left[A_i^2(\phi)g_{\mu\nu}, \psi_i\right].
\ee
Here, the notation $A^2_i(\phi)$ captures the fact that the individual matter fields, $\psi_i$, do not necessarily all couple to the {\it same} Jordan-frame metric. 
The first restriction we make is to demand that the model obey the WEP,\footnote{Interactions in the dark sector, {\it i.e.}, a coupling between dark matter and DE, are a further scenario to consider.  Since DE cannot strongly interact with visible matter, this approach violates the WEP.  We will not consider this possibility in depth, but see~\cite{Silvestri:2009hh} for a review of the phenomenology.} this restricts the couplings to be the same, $A_i^2(\phi)\equiv A^2(\phi)$. 
Models with $A^2(\phi) = 1$ satisfy the SEP, and hence are models of DE. 
In these models, the scalar field, $\phi$, affects the motion of matter only through its gravitational influence, it is a decoupled source of stress energy. Cases where $A^2(\phi)$ is some nontrivial function are models of MG. 
In MG models, the force mediated by $\phi$ does not affect all objects universally. 
As an extreme example, consider the motion of some diffuse object compared to that of a black hole. Due to the no-hair theorem~\cite{Bekenstein:1995un}, black holes carry no scalar charge and therefore feel no fifth force from $\phi$. 
However, a more diffuse object like a star or planet will feel such a fifth force, leading to a large violation of the SEP.

It is worth mentioning that although the theories we consider satisfy the WEP at the microscopic level (all matter couples to the same metric), it is nevertheless possible to have apparent violations of the WEP for macroscopic objects~\cite{Hui:2009kc,Hiramatsu:2012xj}. 
This occurs in theories with screening mechanisms; whether or not the mechanisms operate depends mainly upon the masses of the objects involved, and therefore large mass objects can fall at a different rate than light objects. This is essentially because there is not necessarily a trivial relationship between the scalar charge carried by objects and their gravitational potential, as there is in scalar-tensor theories without screening mechanisms, which generally only exhibit violations of the equivalence principle when $\PhiN$ is large~\cite{Hui:2009kc}.

Though the distinction based upon the SEP is theoretically clean and satisfying, in practice it is not very useful phenomenologically. A more pragmatic distinction (relevant for the tests we will discuss) is to call anything which has a fifth force MG, and anything else DE. Note that this phenomenological distinction suffers from the drawback that at this level it is somewhat difficult to draw a clear boundary between MG and DE~\cite{Kunz:2006ca}.

\begin{marginnote}
A flowchart explanation of the distinctions made in the text:

\includegraphics[width=.2\textwidth]{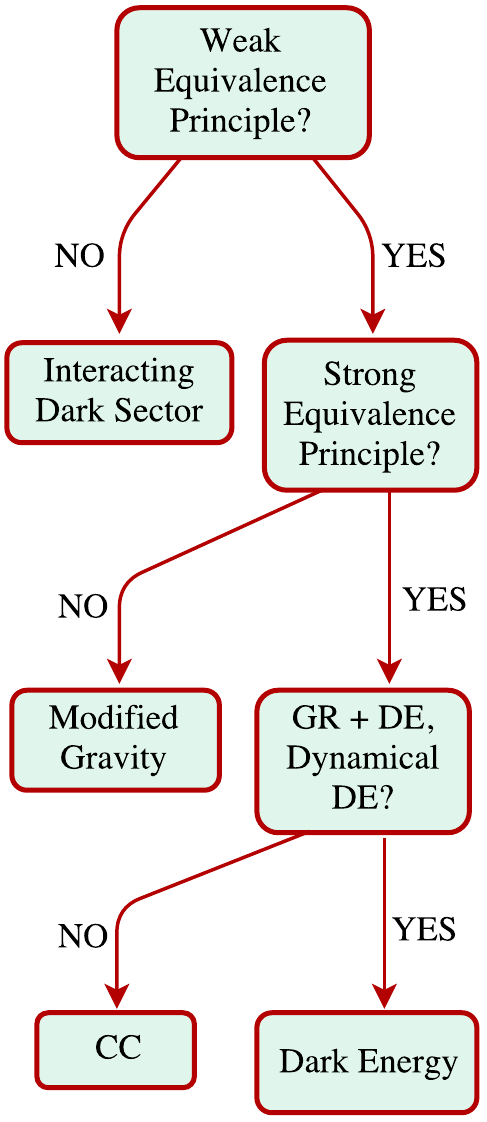}
\label{flowchart1}
\end{marginnote}

\section{COSMOLOGICAL PHENOMENOLOGY}
\label{sec:pheno}

In order to describe the phenomonelogy of DE and MG models, it is
convenient to first consider the unperturbed universe, described by the
scale factor $a(t)$, before turning to the equations governing the
evolution of cosmological perturbations.  As we will see, the 
phenomenology of the latter is much richer, and more likely to yield
insights distinguishing MG and DE.

\subsection{Background Expansion}

In GR, applying the Einstein equations to the Friedmann-Robertson-Walker (FRW) metric yields the
Friedmann equations,
\ba
H^2 =\:& \frac1{3\Mpl^2} \sum_i \rho_i = \frac1{3\Mpl^2} \left[\rhob + \rhoDE\right] \vs
\dot H + H^2 =\:& -\frac2{3\Mpl^2} \sum_i [\bar\rho_i + 3 \bar p_i] 
= -\frac2{3\Mpl^2} [\rhob + (1+3 w) \rhoDE]\,,
\label{eq:Friedmann}
\ea
where we have specialized to the case of a spatially flat
universe and two stress-energy components,
non-relativistic matter (dominated by cold dark matter, CDM) and DE with equation of state $w(t)$, which is the relevant case in the late Universe.  

MG models lead to correspondingly modified Friedmann
equations when applied to a homogeneous cosmology.  As an example,
consider the DGP model (\refsec{massive}), where the first Friedmann equation
is modified to
\be
H^2 \pm \frac{H}{r_c} = \frac1{3\Mpl^2} \rhob\,,
\ee
in the absence of DE.  The negative sign corresponds
to the self-accelerating branch, which admits a de Sitter solution
$H=$~const.~at late times.  However, 
this expansion history is easily mimicked in GR by a suitably chosen quintessence potential that yields $\rhoDE(t)/3\Mpl^2 = H(t)/r_c$.  
This clearly illustrates that, while constraints on the background expansion are
crucial in order to interpret large-scale structure observables, they
do not suffice to cleanly distinguish between MG and DE.

\subsection{Structure Formation with Quintessence}
\label{sec:smDE}

We now turn to the growth of structure in the context of a quintessence
DE model, {\it i.e.}, a canonical light scalar field with effective sound speed
of one.  The CC, {\it i.e.}, the $\Lambda$CDM standard paradigm of cosmology, is included here as a limiting case.  Current constraints already imply that for $z \lsim 1$, the equation of state has to be close to $-1$.  
In these simplest DE models, density and pressure perturbations of the DE component are of order $(1+w) \Phi$, where typical cosmological potential perturbations are of order $\Phi\sim 10^{-4}$.  Further,
anisotropic stress is negligible in these models. 
Thus, DE perturbations have a very small effect on LSS.  

We will continue to assume a spatially flat background and work in the conformal-Newtonian gauge, so that the metric is given by
\be
{\rm d}s^2 = -(1+2\Psi) {\rm d}t^2 + 2 a(t) B_i {\rm d}t {\rm d}\vx^i 
+ a^2(t) \left[ (1-2\Phi) \d_{ij} + \gamma_{ij} \right] {\rm d}\vx^i {\rm d}\vx^j\,.
\label{eq:metriccN}
\ee
Assuming that the scalar potential perturbations $\Psi,\,\Phi$ are much smaller than 1, we can work to linear order in them.  This is accurate
almost everywhere in the universe.  The potential $\Psi$ governs the dynamics of non-relativistic
objects, while the combination $ \Phim \equiv (\Phi+\Psi)/2$ 
determines the null geodesics, {\it i.e.}, light propagation.  One way to see this is
to note that null geodesics are conformally invariant.  Then, consider
a conformal transformation $g_{\mu\nu} \mapsto e^{2\omega} g_{\mu\nu}$.  At linear
order, this corresponds to $\{\Psi,\Phi\} \mapsto \{\Psi + \omega, \Phi - \omega\}$, so that $\Phim$ is the unique linear combination of potentials that is conformally
invariant.  

In~\refeq{metriccN}, $B_i$ is a transverse vector capturing the vector modes, 
while $\gamma_{ij}$ is a transverse-tracefree tensor corresponding to 
gravitational waves (tensor modes).  We will neglect the vector modes
throughout, as they decay and do not propagate.  We will briefly consider tensor modes in \refsec{tensor}.  

\begin{marginnote}
The majority of large-scale structure measurements are on \mentry{subhorizon scales $\bm{k \gg aH}$}.  In this limit, time derivatives, which are of order
$H$ such that $\dot\Phi \sim H \Phi$, can be neglected compared to spatial derivatives.
\end{marginnote}

In GR, the difference between the potentials $\Phi-\Psi$ is sourced
by the anisotropic stress, {\it i.e.}, the trace-free part of $T_{ij}$.  In the models 
considered in this section, this is negligible so we can set $\Phi=\Psi$
(see \refsec{bsmDE} for generalizations).  
On \emph{subhorizon scales} $k \gg a H$,  the 00-component of the
Einstein equations reduces to the Poisson equation,
\be
\nabla^2 \Phi = 4\pi G\, \rhob\, a^2\, \d\,, \quad\mbox{or}\quad
\nabla^2\Phi = \frac32 \Om(a)\, (aH)^2 \d\,,
\label{eq:Poisson}
\ee
where $\d = \rho/\rhob - 1$ is the fractional matter overdensity, and $\Om(a) = \rhob(a)/3 H^2 \Mpl^2$.  For later reference, we define the solution to \refeq{Poisson} as Newtonian potential $\PhiN$.  The continuity and Euler equations for the collisionless DM fluid are
\ba
\dot\d + \frac1a \partial_k \left[(1+\d) \v{v}^k\right] =\:& 0 \vs
\dot{\v{v}}^i + H \v{v}^i + \frac1a \v{v}^k \partial_k \v{v}^i  =\:& - \frac1a \partial^i\Psi\,,
\label{eq:Euler}
\ea
where $\v{v} = a {\rm d} \vx/{\rm d}t$ is the fluid peculiar velocity. 
On large scales, for Fourier modes $k \lsim 0.05 \iMpch$, both
the density perturbation $\d$ and the peculiar velocities $|\v{v}|$
are much less than one, and we can linearize these equations.  
They can then be combined to yield the \emph{linear growth equation}
\be
\ddot\d(\vk,t) + 2 H \dot\d(\vk,t) = -\frac{k^2}{a^2} \Psi(\vk,t) = \frac32 \Om(a) H^2 \d(\vk, t)\,,
\label{eq:lingrowth}
\ee
while the velocities obey $\v{v} = -i a\,\v{k}/k^2 \dot\d$.  Thus, individual
Fourier modes of the density field evolve independently and at the same rate on
all scales.  Furthermore, the evolution of
density and velocity (given initial conditions) are completely controlled by the expansion
history $H(t)$ of the universe.  
Thus, quintessence predicts a definite relation between the observed expansion history and growth of structure, allowing for consistency tests (see \refsec{constests}).  

Generally, for fixed initial conditions
and an approximately constant equation of state $w$, 
the amount of late-time structure monotonically decreases with increasing
$w$ (see the curves for $c_s=1$ in \reffig{Pk-DE}), as the accelerated expansion sets in earlier for less negative $w$.  

The same conclusions remain valid when considering the full
Euler-Poisson system \refeqs{Poisson}{Euler}, and, indeed the exact
Vlasov (collisionless Boltzmann) equation for dark matter:  the 
only impact of quintessence DE is through the background expansion history.  

The nonlinear regime of LSS is frequently described by the \emph{halo model}
reviewed in \cite{cooray/sheth}.  This is motivated by the fact that a significant 
fraction of the dark matter resides in self-gravitating collapsed structures
called halos.  Moreover, the majority of observed LSS tracers,
such as galaxies and clusters, are found in these halos.  
In contrast to the complicated processes governing galaxy formation,
collisionless N-body simulations are able to accurately predict abundance and clustering of halos.  

The mass function of halos, that is, their number density in a logarithmic
mass interval, is described to within $\sim 5$\% (in the context of $\Lambda$CDM \cite{tinker/etal:08}) by the universal form
\be
\frac{{\rm d}n(M,z)}{{\rm d}\ln M} = \frac{\rhob}{M} f(\nu) \left|\frac{{\rm d}\ln\sigma(M,z)}{{\rm d}\ln M}\right|,\quad \nu \equiv \frac{\delta_c(M)}{\sigma(M)}\,,
\label{eq:dndlnM}
\ee
where $\sigma(M,z)$ is the variance of the linear density field at redshift
$z$ when smoothed with a tophat filter of radius $R(M)$ which contains the mass $M$
at the background density $\rhob$, while $f(\nu)$ is in general a free function.  
$\delta_c$ is the initial overdensity of a spherical tophat
overdensity which collapses (reaches radius $R_{\rm th}=0$) at given redshift $z$, extrapolated
from the initial time to redshift $z$ through linear growth.  For quintessence DE, the tophat radius obeys ({\it e.g}., \cite{HPMhalopaper})
\be
\frac{\ddot R_{\rm th}}{R_{\rm th}} = -\frac{4\pi G}3 \left[\rhob + (1+3 w)\rhoDE\right] - \frac{4\pi G}3 \d\rho_m\,,
\label{eq:scollapse}
\ee
where $\d\rho_m = \rho_m(<R_{\rm th}) -\rhob$ is the overdensity in the interior of the tophat.  
\refeq{dndlnM} is inspired by the excursion set approach \cite{Bond/etal:91},
and the well-known Press--Schechter mass function \cite{PS74} is a special
case.  Massive halos correspond to small variances $\sigma(M)$ and hence
to large $\nu$.  In the high-$\nu$ limit, $f(\nu)$ asymptotes to 
$\exp(-q \nu^2)$, $q = \O(1)$, corresponding to exponentially
rare high-mass halos.  The abundance of halos depends on the growth history both through $\sigma(M)$ and $\d_c$, 
although the dependence of $\delta_c$ on cosmology is quite weak.  

The large-scale distribution of halos, described statistically by their
$N$-point functions, can also be derived in this picture.  The most
important statistic on large scales is the two-point function or
power spectrum, $P(k)$.  On scales where linear perturbation theory applies,
the halo two-point function is related to that of matter by $P_h(k,z) = b_1^2(z) P_m(k,z) + C$, where $C$ is a constant corresponding to white noise, and
the linear bias parameter $b_1$ can be derived from the halo mass function
\refeq{dndlnM}.  $b_1$ describes the response of the abundance of halos
to a long-wavelength density perturbation $\d_l$, or equivalently a change in
the background density $\rhob \mapsto \rhob(1+\d_l)$.  This corresponds to 
lowering the threshold $\d_c \mapsto \d_c - \d_l$, so that the linear halo
bias becomes \cite{mo/white:96}
\be
b_1(M,z) = \frac{\rhob}{n(M,z)} \frac{\partial n(M,z)}{\partial\rhob}
= -\frac1{\sigma(M) f(\nu)} \frac{{\rm d}f(\nu)}{{\rm d}\nu}\,.
\ee
Finally, assuming a universal density profile for halos $\rho(r|M,z)$, for
example the NFW profile \cite{NFW} $\rho(r|M,z) \propto [r (1 + r/r_s)^2]^{-1}$, 
the halo model provides a description
of the statistics of the matter density field on nonlinear scales.

\subsection{Beyond Quintessence Dark Energy}
\label{sec:bsmDE}

We now consider three relevant generalizations of the quintessence DE
case considered in the previous section: small speed of sound $c_s = (\delta p/\delta \rho)^{1/2} \ll 1$,
anisotropic stress, and an interacting dark sector.  

\begin{figure}
\includegraphics[width=0.6\textwidth]{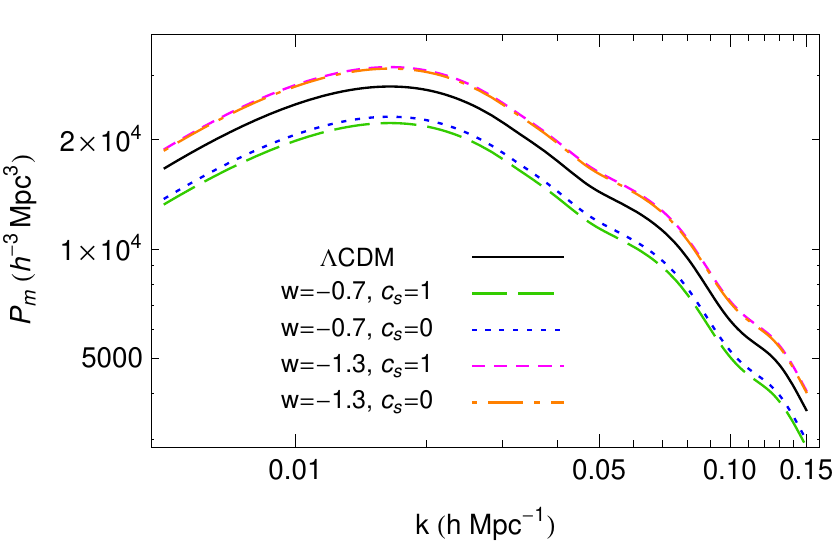}
\caption{Linear matter power spectrum in quintessence DE with different values of equation of state $w$ and sound speed $c_s$.  For the values of 
$w$ shown, clustering DE with vanishing sound speed modifies the matter power
spectrum by a few percent.  This can be probed by weak lensing measurements of the CMB and galaxies (\refsec{CMB}--\ref{sec:nonlinearobs}). Reproduced from~\cite{creminelli/etal:09}.}\label{fig:Pk-DE}
\end{figure}

DE density perturbations are negligible on
scales that are much smaller than the sound horizon of the DE,
$R_{s,\rm DE} \simeq c_s H^{-1}$ in physical units.  Let us now consider the opposite
case, {\it i.e.}, what happens on scales much greater than $R_{s,\rm DE}$.  On those
scales, the pressure perturbations of the DE component are negligible
compared to its density perturbations.  This means that the DE fluid moves on the same trajectories
(geodesics) as the pressureless CDM component.  Hence, density perturbations
grow on small scales in both components, although the DE perturbations are
suppressed by $1+w$ (on intermediate scales $c_s k/aH \sim 1$, DE pressure perturbations need to be taken into account).  Specifically, the Euler--Poisson system is augmented
by the continuity equation for the DE component and becomes
\ba 
\dot\d_{\rm DE} - 3 H w\, \d_{\rm DE} + \frac1a \partial_k \left[  (1+w + \d)\v{v}^k \right] = 0\,; \quad\quad
\nabla^2 \Phi = 4\pi G (\rhob \d + \rhoDE \d_{\rm DE})\,.
\nonumber
\ea 
It is also possible to extend the spherical collapse \refeq{scollapse}
to this case \cite{creminelli/etal:09}:
\ba
\frac{\ddot R_{\rm th}}{R_{\rm th}} =\:& -\frac{4\pi G}3 \left[\rhob + (1+3 w)\rhoDE\right] - \frac{4\pi G}3 (\d\rho + \d\rho_{\rm DE})\, ;
\label{eq:scollapsecs}\\
\dot\rho_{\rm DE}(< R_{\rm th}) =\:& - 3 \frac{\dot R_{\rm th}}{R_{\rm th}} \left[\rho_{\rm DE}(<R_{\rm th}) + w \rhoDE\right]\,.
\ea
Again we see that if $1+w\ll 1$, then initially small perturbations in the DE density will stay small.  
\reffig{Pk-DE} illustrates the effects of $c_s=0$ DE for the linear matter power spectrum.  For $w=-0.9$, the effect on the matter power
spectrum is of order 1\% at $z=0$ (and smaller at higher redshifts).  
The halo abundance on the other hand, calculated using \refeq{dndlnM} with the modified 
$\sigma(M)$ and $\d_c$, is larger by 5-10\% at the very high mass end 
for $c_s=0$ as compared to $c_s=1$ \cite{creminelli/etal:09}.  
As noted by \cite{deputter/huterer/linder}, the quantitative effect of
small DE sound speed can be increased if there is a significant ``early DE'' component.  

We now turn to the case of a DE component with scalar anisotropic
stress $\Pi_{ij}^{\rm DE} \equiv T_{ij}^{\rm DE} - \d_{ij} T^{\rm DE}_{kl}\d^{kl}/3$.  $\Pi_{ij}^{\rm DE}$ sources a difference between the two metric potentials following
\be
\left[\partial_i\partial_j - \frac13 \delta_{ij} \nabla^2 \right] (\Phi-\Psi) 
= 8\pi G\,a^2\, \Pi_{ij}^{\rm DE}\,.
\ee
The fact that $\Phi\neq\Psi$ is relevant to the cause of distinguishing between MG and
DE, since the difference between $\Phi$ and $\Psi$ is otherwise a 
distinctive feature of MG (see the following section).  DE models that involve a single scalar field
do not lead to anisotropic stress, while models involving vector fields
do in general \cite{picon:04,wei/cai:06}.  
Ref.~\cite{koivisto/mota,mota/etal:07} investigated the impact on structure
formation of DE with anisotropic stress based on the effective 
parameterization of \cite{Hu:98} via a viscous parameter $c_{\rm vis}$.  
They show that the effect of DE anisotropic stress observable via
$\Phi-\Psi$ is only significant
for horizon-scale perturbations.  On the other hand, MG generally leads
to a significant $\Phi-\Psi$ on all scales, so that one can still hope
to effectively distinguish between MG and DE via this probe.  

Finally, one can also couple the DE component to the non-baryonic
components, such as CDM or neutrinos \cite{farrar/peebles,amendola/baldi/wetterich}.  
Following our distinction in \refsec{dab}, these are not modified
gravity models, since the fifth force does not obey the WEP.  The absence of coupling to baryons and radiation lets these
models evade local tests, and hence the effects on the large-scale
structure can become quite significant \cite{baldi/etal:10}.  Interestingly, a
generic signature of this model is a different clustering amplitude (bias)
of baryons relative to the dark matter on large scales, which can
be used to place constraints on this type of model.

\subsection{Structure Formation in Modified Gravity} \label{sec:strMG}

In this section, we briefly review the rich phenomenology of structure growth in MG, separately for the linear and nonlinear regimes.

\subsubsection{Scalar perturbations on linear scales} \label{sec:linear}

We begin with structure formation in MG theories on sufficiently large scales,
where linear perturbation theory applies, and assume that
stress-energy perturbations are only due to a CDM component.  
It is convenient to work in Fourier space, where individual modes evolve independently at linear order.  Then, given that
the spacetime is described by two potentials $\Phi(\vk, t),\,\Psi(\vk,t)$,
while the matter sector is characterized completely by $\d(\vk,t)$ 
(the velocity is determined via the continuity equation),
any theory of gravity can be described in this regime by two free functions
of $\vk$ and $t$ which parameterize the relation between the three fields
$\Phi,\,\Psi,\,\d$.  

On superhorizon scales $k \ll aH$ and for adiabatic perturbations, further constraints are posed by diffeomorphism invariance \cite{bertschinger:06,Hu:2007pj,gleyzes/etal:15}. 
Specifically, a superhorizon adiabatic mode behaves like a separate, curved universe, so that its evolution is constrained by the solution to the background expansion in the given theory.  Ref.~\cite{bertschinger:06} derived
\be
\ddot \Phi - \frac{\ddot H}{\dot H} \dot\Phi + H \dot\Psi
+ \left(2 \frac{\dot H}{H} - \frac{\ddot H}{\dot H}\right) H \Psi = 0\,,
\label{eq:ddotPhiSH}
\ee
which, given one function of time specifying the relation between $\Phi$ and $\Psi$ for $k\ll aH$,
determines the superhorizon evolution of the potentials.  

\begin{marginnote}
\entry{Parametrized post-Friedmann (PPF) approach}{On linear scales and for adiabatic initial conditions, 
any theory of gravity can be parameterized by two free functions
of scale and time [\refeq{MGparam}], with an additional 
constraint on superhorizon scales [\refeq{ddotPhiSH}].}  
\end{marginnote}

We now turn again to the subhorizon limit, and assume adiabatic initial conditions.  In the absence of preferred directions, we can parameterize the relation between $\d$, and $\Phi,\Psi$ via
\ba
-k^2 \Psi(\vk,t) =\:& \frac32 \Om(a) (a H)^2 \,\mu(k,t) \d \vs
\frac{\Phi(\vk,t)}{\Psi(\vk,t)} =\:& \gamma(k,t)\,.
\label{eq:MGparam}
\ea
These equations reduce to GR for $\mu=\gamma=1$.  Given adiabatic initial conditions and assuming the WEP, $\Phi,\,\Psi,\,\d$ are 
at linear order all related
to the initial conditions ({\it i.e.}, a single stochastic variable) through transfer functions $T_\Phi,\,T_\Psi,\,T_\d(k,t)$, so that $\mu \sim (k/aH)^2 T_\Psi/T_\d$
while $\gamma = T_\Phi/T_\Psi$.  Moreover, for a local
four-dimensional theory of gravity, $T_i$ are functions of $k^2$ only, so
that $\mu$ and $\gamma$ reduce to rational functions of $k^2$.  If
we further assume that no higher than second derivatives appear in the
equations for $\Phi,\,\Psi$, then $\mu$ and $\gamma$ are completely described by five functions
of time only \cite{silvestri/pogosian/buny}
\be
\mu(k,t) = \frac{1 + p_3(t)k^2}{p_4(t) + p_5(t) k^2} ; \quad
\gamma(k,t) = \frac{p_1(t) + p_2(t) k^2}{1 + p_3(t)k^2} \,. \quad
\label{eq:mugamma}
\ee
Note that higher powers of $k$ can appear in the equations for $\Phi,\,\Psi$
after integrating out the additional d.o.f., even if the fundamental equations
are all second order in derivatives.  
Using suitable interpolation, these subhorizon limit results can be connected to superhorizon scales in a parameterization that enforces \refeq{ddotPhiSH} \cite{Hu:2007pj,bertschinger/zukin,baker/etal:13}, which further reduces the number of free functions.   Finally, Refs.~\cite{Gleyzes:2013ooa,bellini/sawicki} found that four free functions of time, along with a free background expansion history, completely describe the linear growth of perturbations for the Horndeski lagrangian \refeq{Horndeski}.

While different equivalent parameterizations exist,
the specific choice of $\mu$ and $\gamma$ adopted here
is motivated by the fact that $\mu$ can be directly constrained
by the growth of structure, as $\Psi$ is the potential governing
the motion of matter.  From \refeq{lingrowth}, the growth equation is
modified to
\be
\ddot\d(\vk,t) + 2 H \dot\d(\vk,t) = \frac32 \Om(a) H^2 \mu(k,t)\d(\vk,t)\,.
\label{eq:lingrowthmu}
\ee
That is, in MG, the linear growth is in general scale-($k$-)dependent,
unlike the quintessence case or the case for DE with $c_s = 0$.  Measuring the growth history as a function
of $k$ and $t$ in principle allows for a measurement of $\mu(k,t)$.

On the other hand, $\gamma$, also referred to
as \emph{gravitational slip}, quantifies the departure between the
two spacetime potentials.  
The propagation of photons is, in the coordinate frame \refeq{metriccN},
governed by the combination of potentials 
\be
\Phim = \frac12 (\Psi+\Phi) = \frac12 (1+\gamma) \Psi\,.  
\label{eq:Phim}
\ee
This is in exact analogy to the $\gamma_{\rm PPN}$ parameter in the parameterized
post-Newtonian (PPN) framework 
\cite{Will:2014kxa}, which is similarly constrained
by combining photon propagation (Shapiro delay) with dynamics (the Earth's orbit).  The fact that Solar System tests constrain $|\gamma_{\rm PPN}-1| \lsim 10^{-5}$ 
clearly shows that a consistent modification of gravity in cosmology
has to be scale- or environment-dependent.  
Gravitational lensing observables in cosmology are then approximately related to the matter
density through the combination $(1+\gamma) \mu$ of the MG
parameterization (see \refsec{tests}).

\subsubsection{Scalar perturbations on nonlinear scales} \label{sec:nonlinear}

The description of nonlinear structure formation in the context of MG
is complicated by the necessity of screening mechanisms.  That is,
while we can in principle extend the parameterization \refeq{MGparam}
to small, nonlinear scales, this will in general violate Solar
System constraints on gravity, which tightly constrain $\mu$ and $\gamma$.  
The most interesting MG models can circumvent these
constraints by employing nonlinear screening mechanisms.  By 
definition, they are not captured by a linear parameterization 
of the form \refeq{MGparam}.  For this reason, cosmological constraints
on MG from nonlinear scales are generically model-dependent.  
In this context, it is useful to classify models by their screening mechanism
(\refsec{screening}).  

In models of the chameleon or symmetron/dilaton type, a certain depth of the gravitational potential
is necessary to pull the field away from its background value $\bar\varphi$
and activate the screening mechanism.  Specifically, if the
Newtonian potential $\PhiN$, which is the solution of \refeq{Poisson},
satisfies $-\PhiN \gsim C \bar\varphi/\Mpl$, then $\varphi$ becomes locally
suppressed compared to $\bar\varphi$, and the deep potential region
is screened.  The constant $C$ depends on the model,
but is typically of order unity; for $f(R)$ gravity, $C=3/2$.  In order for 
the Solar System to be screened, $C \bar\varphi/\Mpl \lsim 10^{-5}$, which 
can be shown to put an upper limit on the mass $\bar m_{\varphi} \gg H$ of the field in the background \cite{Wang:2012kj}.  
The result is that viable models with this screening mechanism only modify
structure formation on scales below $\sim 30$~Mpc.  

\begin{figure}
\centering
\begin{minipage}{.5\textwidth}
  \centering
\includegraphics[width=\textwidth]{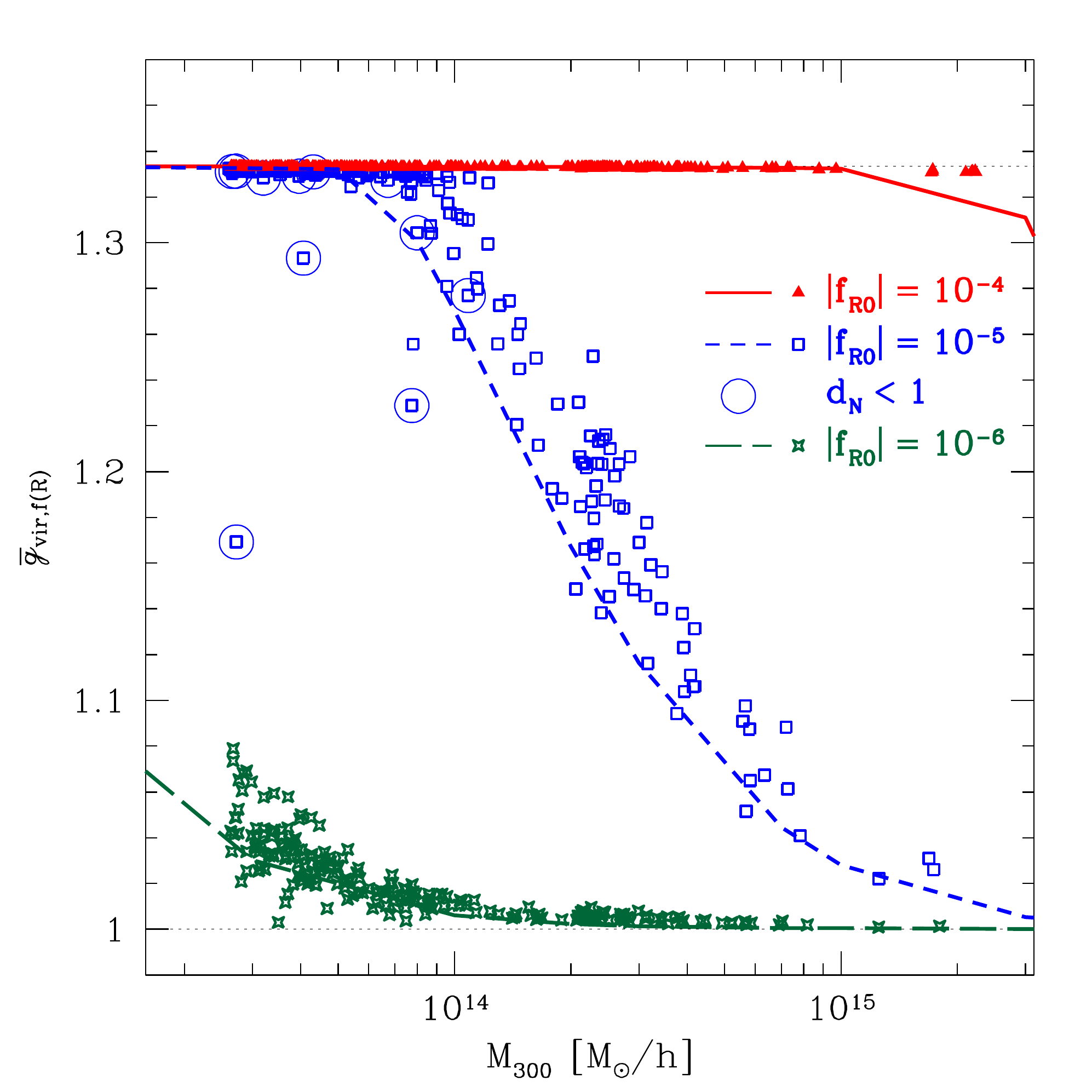}
\end{minipage}%
\begin{minipage}{.5\textwidth}
  \centering
\includegraphics[width=\textwidth]{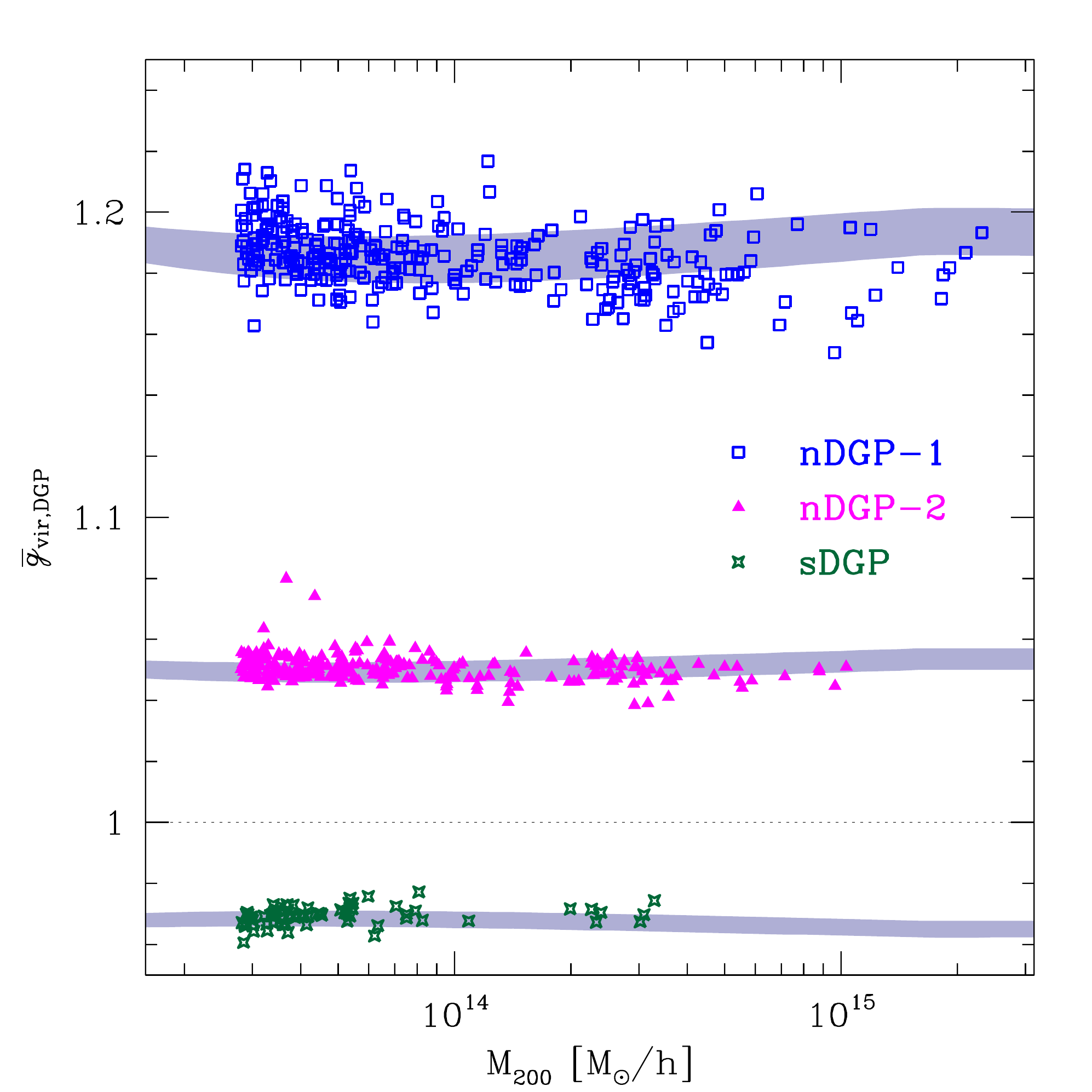}
\end{minipage}%
\caption{Mean mass-weighted gravitational acceleration (relative to GR) 
$\bar{\mathscr{g}}$ within dark matter halos, as measured in MG simulations for $f(R)$ (left panel), corresponding to chameleon screening,
and DGP (right panel), showing Vainshtein screening.  The ratio of dynamical to lensing mass of these halos is given by $M_{\rm D}/M_{\rm L} = \bar{\mathscr{g}}^{3/5}$ (\refsec{targeted}).  This can be probed by comparing dynamical
mass estimates of galaxy clusters with their gravitational lensing signal
(\refsec{nonlinearobs}).  
The $f(R)$ model is of Hu-Sawicki type 
\cite{Hu:2007nk} [\refeq{husawickifr}].  ``sDGP'' corresponds
to the self-accelerating branch where gravity is weakened, while the ``nDGP'' models show normal-branch DGP with a quintessence component added to produce
an expansion history identical to $\Lambda$CDM 
[\cite{2009PhRvD..80l3003S}; with $r_c = 500,\,3000$~Mpc for nDGP-1, nDGP-2, respectively].  
Reproduced from \cite{2010PhRvD..81j3002S}.}\label{fig:screening}
\end{figure}

This screening behavior further means that there will be
a characteristic threshold halo mass $M_{\rm scr}(\bar\varphi)$, with
halos above this mass being screened while halos below this threshold
are unscreened (since halo profiles are nearly universal, there is a well-defined mapping between central potential and halo mass).  This transition effect can be clearly seen
in N-body simulations of chameleon-type models;  the left panel of \reffig{screening} shows the gravitational acceleration within dark matter halos in $f(R)$ simulations
which consistently include the chameleon mechanism \cite{2008PhRvD..78l3523O}.  
Halos with a potential depth larger than the field value $-\PhiN \gsim 2|f_{R0}|/3$
become screened, while lower mass halos are unscreened.  The lines
show the expectations based on a simple spherical model of the halos
\cite{2010PhRvD..81j3002S}.  The circled points show halos that
are screened due to the potential of a massive halo in the vicinity,
rather than their own potential well.  Since unscreened halos accrete mass at
a rate higher than in GR, this transition effect is also manifested in the
halo mass function \cite{HPMhalopaper}.  
The halo of our own galaxy has to be screened in order for a chameleon/symmetron
model to satisfy Solar System tests, so that these models can only be expected
to show MG effects in lower mass halos.  This effect motivates the search
for MG effects in nearby dwarf galaxies (\refsec{submpc}).  

We now turn to the Vainshtein screening mechanism; for simplicity we restrict
to the cubic Galileon interaction term [\refeq{cubicG}].  In the case of spherical symmetry, the equation of motion in the subhorizon limit can be integrated once to yield 
\be
\frac{{\rm d}\varphi}{{\rm d}r} \propto \frac{M(<r)}{\Mpl\, r^2} g\left(\frac{r}{r_*(r)}\right) 
\quad\mbox{where}\quad
g(\xi) = \xi^3 \left(\sqrt{1+\xi^{-3}} - 1\right)\,,
\label{eq:dphdr}
\ee
where $M(<r)$ is the mass (over the background density) enclosed within $r$.  The scale
$r_*(r) \propto [M(<r) / \Mpl \Lambda^3]^{1/3}$  is the $r$-dependent \textit{Vainshtein radius} (this scaling also holds for higher order galileon interactions). 
In the case of the DGP model, it is given by 
\be
r_*(r) = \left[\frac{16 M(<r) r_c^2}{9 \Mpl^2 \beta^2}\right]^{1/3}\,,\quad
\beta = 1 \pm 2 H r_c \left[1 + \frac{\dot H}{3 H^2}\right]\,.
\label{eq:rstar}
\ee
We see that ${\rm d}\varphi/{\rm d}r$ is suppressed compared to the Newtonian gradient
$\Mpl\,{\rm d}\PhiN/{\rm d}r$ for $r/r_* \ll 1$, where the quantity $(r/r_*)^3$ is directly
proportional to the inverse of the interior density $\rho^{-1}(<r)$.  
That is, Vainshtein screening occurs at fixed interior density, 
in contrast to the chameleon type which occurs at fixed mass.  The
threshold density for screening is of order the background matter density 
today for natural model parameters (in the case of DGP, $r_c \sim H_0^{-1}$).  
For nonlinear LSS, this leads to a qualitatively different behavior compared to the chameleon case (right panel of \reffig{screening}):   halos of all masses are screened within a fixed fraction of their scale radius.  Thus, Solar System constraints do not force us to look to certain types of objects for interesting signatures of Vainshtein-type models.

\subsubsection{Tensor perturbations}
\label{sec:tensor}

Besides modifying the scalar perturbations, MG in general also alters the propagation of tensor modes.
More specifically, a running of the gravitational coupling causes a change in the decay of the gravitational wave amplitude in an expanding universe.   Further, the $G_4$ and $G_5$ terms of a Horndeski scalar-tensor theory alter the propagation speed of tensor modes~\cite{Gleyzes:2013ooa,saltas:14}.

\section{COSMOLOGICAL TESTS}
\label{sec:tests}

In this section, we provide an overview of the observables and experimental methods that are used to probe gravity and DE on cosmological scales.

\begin{figure}[t]
\includegraphics[width=3in]{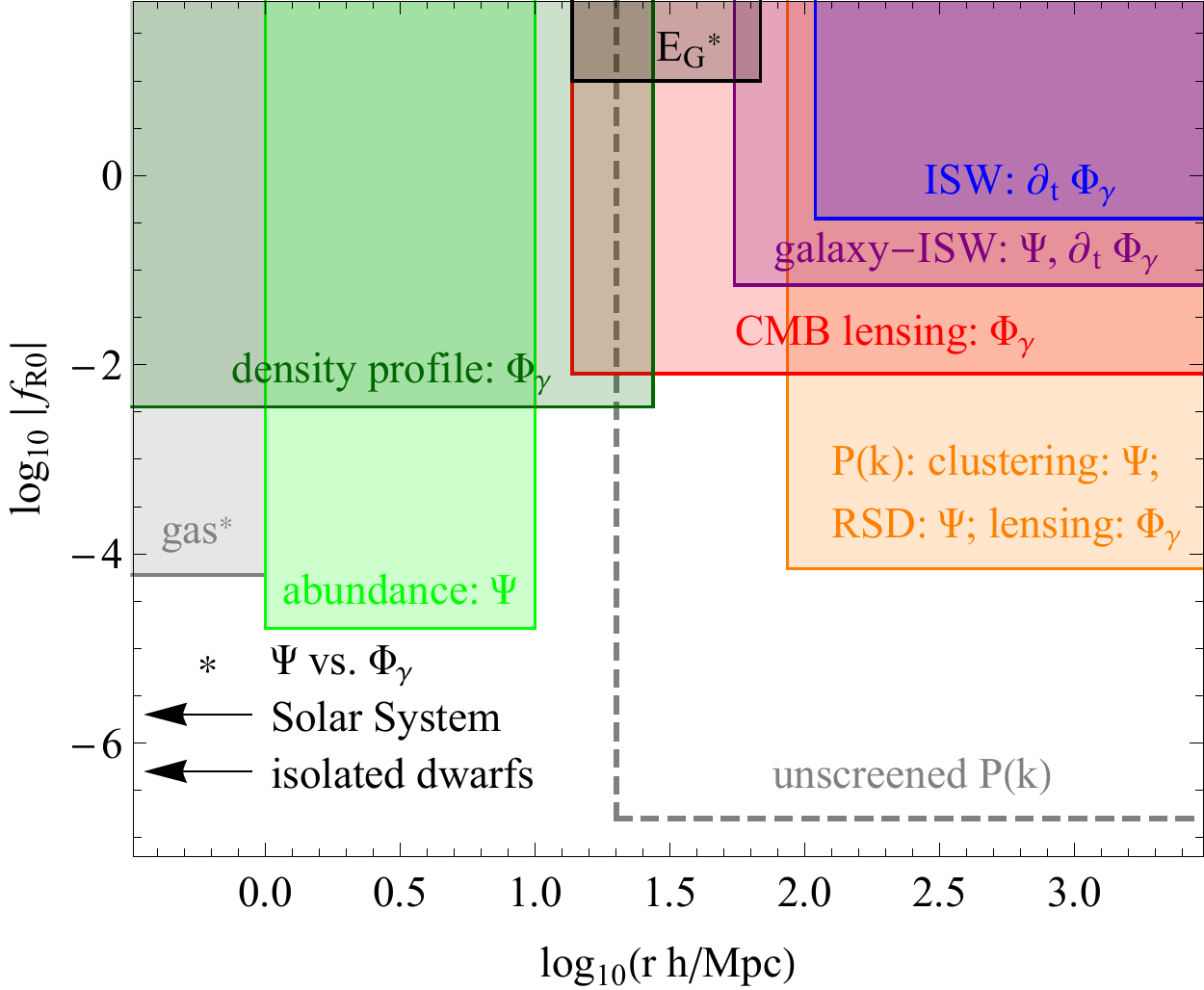}
\caption{Current and forecasted constraints on $f(R)$ gravity, in particular the parameter $f_{R0}$ of the Hu-Sawicki model [\refeq{husawickifr}], and potentials tested (reproduced from~\cite{lombriser:11}).
Linear constraints: ISW and galaxy-ISW cross correlations~\cite{lombriser:10}; CMB lensing~\cite{hu:13}, the $E_G$ probe~\cite{reyes:10}, and power spectrum constraints from galaxy clustering, redshift-space distortions, and weak gravitational lensing~\cite{hojjati:15}.
Nonlinear constraints: cluster abundance~\cite{schmidt:09,lombriser:10,cataneo:14}; cluster density profiles~\cite{lombriser:11}; and the comparison of cluster gas and weak lensing measurements~\cite{terukina:13}.
Also indicated are the currently tightest constraints on $|f_{R0}|$ from Solar System tests~\cite{Hu:2007nk} and distance indicators in isolated dwarf galaxies~\cite{jain:12}.
Future cosmological surveys measuring the power spectrum $P(k<0.3~{\rm h/Mpc})$ in unscreened regions at the $1\%$ level will be able to outperform the astrophysical constraints~\cite{lombriser:15a}.
}
\label{fig:fRconstraints}
\end{figure}

\subsection{Observables} \label{sec:observables}

\noindent MG and dynamical DE models in general change the background expansion history of our Universe with respect to $\Lambda$CDM.
The predicted deviation can be tested by measuring distances to astronomical objects and cosmic ``standard rulers.''  We refer to these measurements
as \emph{geometric probes}.  
Comparing, for instance, the magnitudes of high-redshift to low-redshift Type Ia supernovae, serving as standard candles, yields a measurement of the evolution of the luminosity distance $d_L(z)=(1+z)r(z)$, where $r(z)\equiv(1+z)d_A(z)$ is the comoving angular diameter distance.  
For a spatially flat universe, $r(z)$ is simply the comoving radial distance $\chi(z)=\int_0^z dz'/H(z')$.  
This is a relative distance measurement, although calibration with the
low-redshift ``distance ladder'' also yields an absolute distance measurement.  
Observations of Cepheid stars in supernovae host galaxies yield $H_0=(73.8\pm2.4)$~km/s/Mpc~\cite{riess:11} for the
Hubble constant today.  
Measurements of the acoustic peaks in the cosmic microwave background (CMB), 
and their imprint on large-scale structure---the baryon acoustic oscillations (BAO) feature observed in galaxy clustering---serve as a standard ruler, and provide complementary information about the absolute distance scale to $z \gsim 0.1$.  
These geometric probes together with recent Planck measurements constrain a constant DE equation of state in a spatially flat universe at $w=-1.006\pm0.045$~\cite{ade:15}, consistent with the CC in $\Lambda$CDM.

When DE is evolving or gravity is modified, this also leaves imprints in the formation of the large-scale structure as can be seen from \refeq{lingrowth}. MG models can match the $\Lambda$CDM expansion history at an observationally indistinguishable level while in comparison, the growth of cosmic structure may still be significantly modified, rendering the latter a vital probe for MG. Nonetheless, geometric probes provide important constraints on cosmological parameters which limit degeneracies in growth of structure constraints.
In the following, we review some of the important observables of cosmological structures that are used to test MG and DE.
In \reffig{fRconstraints} we summarize some of the cosmological constraints that have been inferred on $f(R)$ gravity, which shall serve here as a representative for typical MG models that has also been particularly well tested.
We show the range of scales covered by different cosmological observables, and also indicate which of the metric potentials in \refeq{metriccN} they probe.
In Table~\ref{tab:constraints}, we furthermore provide a summary of the current observational constraints on the modified gravity models discussed in Sec.~\ref{sec:theory}.

\begin{table}
\tabcolsep7.5pt
\caption{Current observational constraints on modified gravity models.}\label{tab:constraints}
\label{tab1}
\begin{center}
\begin{tabular}{@{}l|c|c|c|c@{}}
\hline
Screening & Representative & Linear cosmology & Nonlinear & Sub-Mpc \\
mechanism & model &  & cosmology &  \\
\hline
Chameleon & $f(R)$ gravity$^{\rm a)}$ & $|f_{R0}|\lsim10^{-4}$ & $|f_{R0}|\lsim10^{-5}$ & $|f_{R0}|\lsim(10^{-7}-10^{-6})$ \\
Symmetron & \refeq{Vsymmetron}$^{\rm b)}$ & --- & --- & $\chi\lsim(10^{-7}-10^{-6})$\\
Vainshtein & nDGP$^{\rm c)}$ & $H_0r_c\gtrsim(0.1-1)$ & $H_0r_c\gtrsim0.1$ & $H_0r_c\gtrsim10^{-4}$ \\
Vainshtein & cubic galileon$^{\rm d)}$ & {\rm incompatible ISW} & {incomp.~voids} & {\rm compatible} \\
\hline
\end{tabular}
\end{center}
\begin{tabnote}
$^{\rm a)}$\cite{terukina:13,Hu:2007nk,jain:12,cataneo:14,hojjati:15};
$^{\rm b)}$\cite{Hinterbichler:2010es,jain:12}, for $\chi \equiv M^2/(2M_{\rm Pl}^2)=\mu/(2gM_{\rm Pl}\sqrt{\Lambda})$ and coupling set to $g=1$;  note that cosmology constraints on this model have not been derived so far;
$^{\rm c)}$\cite{lombriser:09,raccanelli:12,xu:13,wyman:10,Clifton:2011jh};
$^{\rm d)}$\cite{barreira:14,barreira:15,Nicolis:2008in}, note that the quartic and quintic galileons are tightly constrained by measurements of the speed of gravitational waves (Sec.~\ref{sec:nonlinearobs}).
\end{tabnote}
\end{table}

\subsubsection{Cosmic microwave background}
\label{sec:CMB}

The CMB radiation is the primary observable for obtaining information about our cosmos.  Of particular importance for DE and MG are constraints on the amplitude and adiabaticity of initial perturbations.  
However, late-time modifications manifest themselves in the CMB temperature and polarization only via secondary anisotropies.
The presence of DE or a CC gives rise to the integrated Sachs-Wolfe (ISW)~\cite{sachs:67} effect, a fluctuation in the temperature field at the largest scales due to the change in energy of the CMB photons $\propto\dot\Phi_\gamma$ when traversing the evolving metric potentials.  This effect is modified in alternative theories of gravity or dynamical DE as can be seen from \refeq{MGparam}.  
A related secondary anisotropy in the CMB is due to the depths of the potentials $\Phim$ which determine the weak gravitational lensing (WL) of the CMB photons by the cosmological structure between the observer and the last-scattering surface.
The directions of the observed CMB photons are displaced by the lensing deflection angle $\v{d}$. The effect is often described by the convergence field $\kappa = \bm{\nabla}_{\v{\theta}} \cdot\v{d}$, where $\bm{\nabla}_{\v{\theta}}$ is the gradient operator on the sphere.  
On smaller angular scales, the Sunyaev-Zel'dovich (SZ) effect, caused by the energy gain of CMB photons by collisions with electrons in the post-reionization universe, provides a complementary signature that probes gravitationally induced motion of matter, and hence the dynamical potential $\Psi$ rather than the lensing potential $\Phim$.  The fairly small \emph{kinetic} SZ effect~\cite{sunyaev:80a} measures the line-of-sight component of the large-scale electron momentum density.  The larger \emph{thermal} SZ effect~\cite{sunyaev:80b} is induced by virialized structures (\refsec{nonlinearobs}).  
In order to extract these late-time signatures, it can also be useful to consider their cross correlation with the foreground structure.

\begin{marginnote}
\entry{CMB}{cosmic microwave background}
\entry{BAO}{baryon acoustic oscillations}
\entry{ISW}{integrated Sachs-Wolfe effect}
\entry{SZ}{Sunyaev-Zel'dovich effect}
\entry{WL}{Weak gravitational lensing}
\entry{RSD}{redshift-space distortions}
\end{marginnote}

At linear order, the ISW temperature fluctuation, the WL convergence, and galaxy density perturbation are given as functions of the position on the sky $\v{\hat{\theta}}$ by
\ba
\Delta T_{\rm ISW}(\v{\hat{\theta}}) & = -2\int_0^{\chi_*} {\rm d}\chi \frac{{\rm d}t}{{\rm d}\chi \frac{\partial}{\partial t}}\Phim(\chi\v{\hat{\theta}},\chi)\,, \label{eq:dTISW} \\
 \kappa(\v{\hat{\theta}}) & = \int_0^{\chi_*} {\rm d}\chi \frac{\chi_*-\chi}{\chi_*\chi}\nabla^2_{\v{\theta}}\Phim(\chi\v{\hat{\theta}},\chi)\,, 
\label{eq:kappa}\\
 g(\v{\hat{\theta}}) & = \int_0^{\chi_*} {\rm d}\chi \frac{{\rm d}z}{{\rm d}\chi} \, b(z) \Pi(z) \delta(\chi\v{\hat{\theta}},\chi)\,,
\ea
respectively, where $\chi_*$ denotes the conformal radial distance to the last-scattering surface, the galaxy bias $b$ is scale-independent but dependent on redshift, and $\Pi$ is a normalized selection function.  
For multipoles $\ell\gsim 10$, one can employ the subhorizon limit and Limber approximation to write the angular cross-power spectra of these quantities as 
\ba
 C_{\ell}^{XY} & = \int {\rm d}z \frac{H(z)}{\chi^2(z)} \left.\left[ F_X(k,z) \, F_Y(k,z) \, P(k) \right]\right|_{k=\ell/\chi(z)} \,,
\quad\mbox{where}
\label{eq:crosscorrelation} \\
 F_{\rm ISW}(k, z) & \equiv T_{\rm CMB} \frac{3H_0^2\Omega_{m,0}}{k^2} \frac{\partial}{\partial z} G(k, z)  \vs 
 F_g(k, z) & \equiv b(z) \Pi(z) D(k,z) \vs 
 F_{\kappa}(k, z) & \equiv \frac{3H_0^2\Omega_{m,0}}{2H(z)} \frac{\chi(\chi_*-\chi)}{\chi_*} G(k, z)\,, \quad\mbox{and} \vs
G(k, z) & \equiv \frac12 [1 + \gamma(k,z)] \mu(k,z) (1+z) D(k, z)
\,. \nonumber
\ea
Here we have used \refeq{MGparam}, defined $D(z,k)\equiv \delta(z,k)/\delta(0,k)$ as the scale-dependent growth factor, and denoted $P(k) \equiv P(k,z=0)$ as the matter power spectrum today.

\subsubsection{Large-scale structure}
\label{sec:nonlinearobs} 

The bulk of cosmological information on dynamical DE and MG is contained in the three-dimensional large-scale structure at low redshifts $z \lsim 5$.  
Galaxy shape correlations probe the lensing potential $\Phim$ via WL, specifically the convergence $\kappa$  [\refeq{kappa}, with $\chi_*$ replaced with the comoving distance to the source galaxies].  Similarly,
angular correlations of galaxies in redshift slices probe the matter
power spectrum on large scales [\refeq{crosscorrelation}].  Further, the 3-dimensional power spectrum of galaxies with spectroscopic redshifts traces the underlying distribution of matter on large scales ($k \ll 0.1 \iMpch$) via
\be
P_g(\vk,z) = b^2(z) P(k,z) + 2 b(z)\, \hat k_\parallel^2 P_{\d\theta}(k,z) + \hat k_\parallel^4 P_{\theta\theta}(k,z)\,,
\label{eq:Pg}
\ee
where $\hat k_\parallel$ is the cosine of $\vk$ with the line-of-sight.  The peculiar
motion of galaxies leads to the terms involving the cross- and auto-correlation
of the matter velocity divergence $\theta$ [redshift-space distortions (RSD)~\cite{kaiser:87}],
which can be used to measure the growth rate of structure $f \equiv d\ln D/d\ln a$.    
Note that the galaxy bias $b$ cannot be predicted from first principles, and must be constrained using lensing or marginalized over (see \refsec{targeted}).

\begin{marginnote}
\mentry{The bulk of information in large-scale structure is on nonlinear scales.}  
Upcoming surveys will measure statistics of galaxy counts
and lensing with high signal-to-noise out to wavenumbers $k > 1 \iMpch$,
while structure becomes nonlinear at $k \gsim 0.1 \iMpch$ (depending on redshift).
\end{marginnote}

The statistics of matter and $\Phim$ can be measured well into the nonlinear regime through galaxy clustering and WL, where \refeqs{crosscorrelation}{Pg} are no longer correct.  Nonlinear effects become important on scales of $k\gsim0.1\iMpch$ (depending on redshift), and potentially even larger scales in MG.  
The interpretation of measurements in the nonlinear regime is complicated by nonlinear gravitational evolution, baryonic feedback effects and screening in MG. Lacking a fundamental description of the nonlinear structure for general DE and MG theories, observational tests have therefore been limited to specific models.

A further, fully nonlinear probe is the abundance of galaxy clusters, governed by the halo abundance ${\rm d}n/{\rm d}\ln M$ [\refeq{dndlnM}], which yields tight constraints on the allowed deviations from $\Lambda$CDM~\cite{schmidt:09}
(\reffig{fRconstraints}); their density profiles have also been used to probe MG \cite{lombriser:11}.  Clusters can be identified via their
galaxy content, X-ray emission, or SZ signal.  A crucial ingredient for cosmological constraints is the relation between observable and halo mass.  
This can be determined through gravitational lensing of background galaxies
or the CMB, which yields the \emph{lensing mass} $M_{\rm L}(<r) \propto r^2 \nabla\Phim(r)$.  
The gas temperature $T_{\rm gas}$ and pressure $P_{\rm gas}$ profiles measured in X-rays and SZ (where SZ only measures the latter) yield estimates of the \emph{dynamical mass} $M_{\rm D}(<r) \propto r^2 \nabla\Psi(r)$.  For this, one assumes hydrostatic equilibrium, which in GR yields
\be
 \frac{1}{\rho_{\rm gas}} \frac{{\rm d} P_{\rm gas}(r)}{{\rm d}r} = - \frac{G\,M_{\rm D}(r)}{r^2}, \label{eq:hydrostateq}
\ee
where for thermal pressure $P_{\rm gas}\propto\rho_{\rm gas}T_{\rm gas}\propto P_{\rm e}$.  The dominant systematic uncertainty in $M_{\rm D}$ from these
measurements are due to the poorly known nonthermal pressure components.  
Similarly, the velocity dispersion of galaxies within clusters probes
$M_{\rm D}$, although the interpretation is hampered by uncertainties in the velocity
bias of galaxies.  This uncertainty becomes smaller on larger scales,
where galaxies exhibit coherent inflow motion onto clusters \cite{2012PhRvL.109e1301L,Zu:2013joa}.  

As pointed out in \refsec{strMG}, a difference between $M_{\rm L}$ and $M_{\rm D}$ is generic to MG.
Along with the screening effects discussed in \refsec{nonlinear}, this clearly needs to be taken into account when constraining MG using cluster abundance.
The discrepancy can also be used as probe of MG itself (see \refsec{targeted}).  The screening of MG effects in dense regions motivates the use of abundance, clustering, and profiles of underdense regions (\emph{voids}) as MG probe~\cite{Hui:2009kc}. 

Finally, with the advent of gravitational wave astronomy demonstrated with the first detection from aLIGO \cite{LIGO:2016}, as well as aVIRGO and eLISA in the future, it will become feasible to test the modifications in the propagation of gravitational waves described in \refsec{tensor} (see also \cite{LIGOtests:2016}).  
In particular, a direct comparison of the arrival times between gravitational waves and the electromagnetic signal from a reliably identified counterpart will place tight constraints on deviations between the propagation speed of tensor modes and the speed of light.  Further, tight constraints on the $G_4$ and $G_5$ terms of Horndeski gravity~\cite{jimenez:15} are placed by binary pulsar timing, as these are not screened by the Vainshtein mechanism (assuming no additional screening mechanism is active).

\subsubsection{Sub-megaparsec scales}
\label{sec:submpc}

While small scales lie beyond the scope of this review, it is worth mentioning that besides the Solar System constraints, 
the tightest bounds on chameleon and symmetron modifications are currently inferred from stellar distance indicators (standard candles) in nearby dwarf galaxies residing in a low-density region of space~\cite{jain:12}.   
This test uses the combination of red giant stars which are expected to be
screened, with Cepheids whose pulsating envelopes are still unscreened.  
The difference in screening introduces a systematic deviation between the distances inferred.  
These constraints depend crucially on a correct identification of unscreened environment of these stars, obtained from a reconstruction of the local density field.  

\begin{figure}
\includegraphics[width=0.8\textwidth]{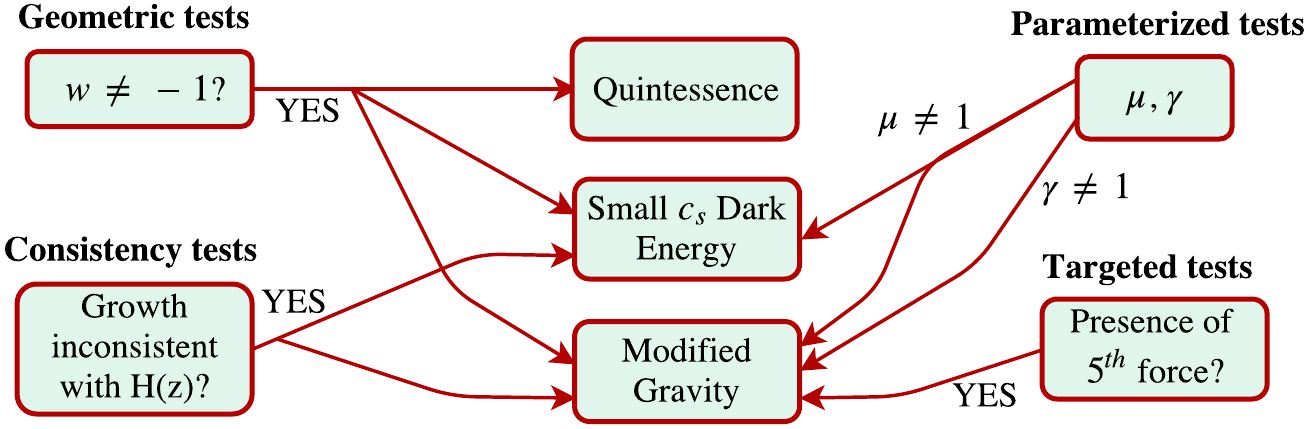}
\caption{Summary of the various types of cosmological tests and their implications we discuss in~\refsec{constests}--\ref{sec:targeted}. Detection of deviations from $\Lambda$CDM phenomenology in these four classes of tests imply different physics beyond the standard cosmology, as indicated by arrows. }
\label{fig:tests}
\end{figure}

\subsection{Consistency Tests: Geometry and Growth}
\label{sec:constests}

The most common approach to obtain cosmological constraints on DE is to
combine various geometric probes of the expansion history and observations
of the growth of structure to obtain joint constraints on the DE equation
of state $w$.  Since only quintessence unambiguously predicts the growth factor $D(a)$ once the expansion history is fixed, these constraints only apply to this type of DE.  Moreover, as discussed in \refsec{quintessence}, a constant $w \neq -1$ 
is not necessarily a good assumption, so that a more general parameterization~\cite{chevallier:00,linder:02}
\be
 w(t)=w_0+w_a[1-a(t)] \label{eq:w0wa}
\ee
is often adopted.  Any observational sign for $w \neq -1$ or $w_a \neq 0$
would be evidence against $\Lambda$CDM.  In order to test for evidence
\emph{beyond quintessence}, Ref.~\cite{Linder:2005in} proposed extending
this parameterization by a parameter $\tilde\gamma$, motivated by the fact 
that the growth factor in a variety of quintessence models satisfies $f = d\ln D(a)/d\ln a = \Om(a)^{\tilde\gamma}$,  
where $\tilde\gamma \approx 0.55$.  Thus, finding $\tilde\gamma \neq 0.55$ would
imply clustering DE, MG, or an interacting dark sector (see also \cite{Ishak:2006}).  
Refs.~\cite{2009PhRvD..79b3004M,2015PhRvD..91f3009R} devised 
more general \emph{consistency tests} between geometry and growth (see \reffig{tests}).  
\begin{marginnote}
Combined constraints on the equation of state $w$ from geometry
and growth \mentry{assume quintessence-type DE}.
\end{marginnote}

While the consistency tests are model-independent, it is generally not true that findings consistent with $\Lambda$CDM rule out any modification of the concordance model, as these parameters do not encompass the entire space of possible DE and MG models.  Further, it is not straightforward to turn a constraint on $\tilde\gamma$ into a constraint on a given MG or non-quintessence DE model, as the parameterization is based on linear subhorizon scales; even if the constraint can be mapped onto a model parameter, it will generally not be an optimal constraint.

\subsection{Parameterized Tests of Gravity}

We can improve several of the disadvantages of the consistency tests discussed above by adopting the more physical parameterization introduced in \refeq{MGparam}, which incorporates physical constraints such as mass and momentum conservation.  Quintessence models correspond to $\mu = \gamma =1$, while clustering DE can be captured by $\gamma=1$ and $\mu \propto 1+w$.  Note however that this parameterization is still restricted to linear scales.

Without further constraints, this parameterization involves two free functions of both time and scale. Ref.~\cite{zhao:09} performed a principal component analysis (PCA), for bins in $k$ and $z$, of $\mu$ and $\gamma$ to determine how well the data can constrain the eigenmodes of $\mu,\,\gamma$ along with redshift bins in $w(a)$.  
They found that the future LSST survey will be able to constrain 
the best 10 eigenmodes in $\mu$ and $\gamma$ to $\sim$3\% and 
$\sim$10\%, respectively.  
While PCA is useful to address the possible modifications in full generality, it may allow for too much freedom to detect a well-defined deviation predicted by a particular clustering DE or MG model.  As discussed in \refsec{linear}, assumptions such as at most second-order derivatives or a Horndeski lagrangian allow us to significantly reduce the freedom in the parameterization, {\it e.g.}, leading to \refeq{mugamma}.  
One might further expect the modifications to scale as the effective DE density $\Omega_{\rm DE} = 1 - \Omega_{\rm K} - \Omega_m$, providing an even more restrictive parameterization of the five functions 
$p_{1-5}(t) = p_{i,0} \Omega_{\rm DE}(t)\Big/\Omega_{\rm DE,0}$.  
Moreover, it can be argued that if the gravitational modifications are to drive cosmic acceleration, then any scale appearing in the equations should be of order the Hubble scale today.  Then, $\mu$ and $\gamma$ are completely determined by one function of time each on subhorizon scales $k\gg aH$.  
Hence, this motivates the restriction of cosmological tests of $\mu$ and $\gamma$ to the two constants $p_{3,0}/p_{5,0}$ and $p_{2,0}/p_{3,0}$.  Note that this last restriction fails to capture $f(R)$ and other chameleon/symmetron models.  

The main caveat to this type of parameterization is that it is limited to linear scales.  While more parameters can be introduced to model a screening-induced suppression of $\mu,\,\gamma$ on nonlinear scales, this clearly does not realistically model the intrinsically nonlinear screening effects.  Finally, even on linear scales, constraints on $\mu,\,\gamma$ are weakened when allowing for other non-standard ingredients, as we will discuss in the following.

\subsection{Cosmic Degeneracies} \label{sec:cosmicdeg}

When comparing theoretical predictions of DE and MG models against cosmological observations, it is important to discriminate the effects from other signatures of new fundamental physics and complex nonlinear processes.
In particular, massive neutrinos or baryonic feedback effects can compensate for the effects of an enhanced growth of structure in the power spectrum or the halo mass function~\cite{baldi:13,puchwein:13}.  Similar effects can be produced by non-Gaussian initial conditions.  
It has also been pointed out that Horndeski scalar-tensor modifications are endowed with sufficient freedom to allow the recovery of the background expansion history and linear large-scale structure of a $\Lambda$CDM universe~\cite{lombriser:15c} (though these models require some finetuning), which limits a fundamental discrimination between MG and DE or a CC based on cosmological structure on large scales.  
Importantly, the modified propagation of tensor modes (\refsec{tensor}) allows these models to be distinguished from $\Lambda$CDM.

\subsection{Targeted Tests} \label{sec:targeted}

In order to avoid known and unknown degeneracies, it is
desirable to devise tests specifically targeting the distinction
between DE and MG that we have introduced in \refsec{dab}.  
More precisely, we would like to devise tests that
constrain a \emph{universally coupled fifth force}.  The most
promising such test uses the generically predicted difference between dynamical ($\Psi$) and lensing ($\Phi_\gamma$) signatures.  In the absence of a fifth
force, this difference could only be sourced through a large anisotropic stress in the
dark sector, which is difficult to obtain (\refsec{bsmDE}).  

On subhorizon linear scales such a test can be performed by combining the cross-correlation between foreground galaxies and lensing, $P_{g(\nabla^2 \Phim)}$ and the galaxy density-velocity cross correlation $P_{g\theta}$ of the same population of galaxies.  The latter can be extracted from the three-dimensional clustering [\refeq{Pg}].
The ratio~\cite{zhang:07}
\be
 E_G \equiv \frac{P_{g(\nabla^2 \Phim)}}{P_{g\theta}}
\simeq \frac{1}{2}(1+\gamma)\mu \frac{\Omega_m}{f}
\label{eq:EG}
\ee
cancels the linear galaxy bias and isolates a combination of the $\mu$ and $\gamma$ parameters that
are induced by fifth forces, providing a robust test of gravity on linear scales.  
Note that $f$ depends on $\mu$ as well.
\begin{marginnote}
There exists a combination of galaxy--lensing and galaxy--velocity cross correlations, \mentry{$\bm{E_G}$}, that is sensitive to a fifth force while canceling the unknown galaxy bias [\refeq{EG}].
\end{marginnote}

On smaller, mildly nonlinear scales, another comparison between dynamics
and lensing is possible by using the infall of galaxies onto massive
clusters \cite{2012PhRvL.109e1301L,Zu:2013joa}.  
WL around clusters constrains the spherically averaged profile
of $\Phim$.  The two-dimensional cross-correlation between galaxies and clusters
on scales $\sim 5-20$~Mpc, which measures a projection of the galaxy phase space,
contains information on the infall velocity and thus $\Psi$.  This probe
goes beyond $E_G$ in that it is not restricted
to linear scales.  However, effects such as a bias between galaxy velocities and those of the
underlying matter distribution need to be carefully controlled.  

On even smaller scales, one can compare the lensing mass $M_{\rm L}$ with the
dynamical mass $M_{\rm D}$ of clusters (measured using X-ray or SZ, \refsec{nonlinearobs}) or galaxies
(measured using stellar velocities).  If $\gamma$ is scale-independent and
in the absence of screening, $M_{\rm D}/M_{\rm L} = (1+\gamma)^{3/5}$ for virialized objects \cite{2010PhRvD..81j3002S}.  For viable models however, taking into account screening is essential \cite{2010PhRvD..81j3002S,terukina:13} (see also \reffig{screening}).

\section{OUTLOOK}
\label{sec:outlook}

We have provided a brief overview of different physical mechanisms to
explain the accelerated expansion of the universe, and introduced
theoretical and phenomenological distinctions between the two scenarios
of DE and MG.  
In the next 15 years, large-scale structure and CMB surveys (AdvACT, eBOSS, DES, DESI,
Euclid, HSC/PFS, LSST, POLARBEAR, SPT-3G, WFIRST and others) have the potential to constrain dynamical 
DE and departures from GR at the few percent level.  This will either rule out a large
swath of the interesting parameter space of DE and MG models, or 
yield another breakthrough in cosmology with the detection of departures
from $\Lambda$CDM.  Thus, this area of cosmology is certain to yield
interesting results in the near future.

\begin{summary}[SUMMARY POINTS]
\begin{enumerate}
\item The weak and strong equivalence principles provide a means to rigorously
distinguish DE, MG, and an interacting dark sector at the theory level (see the flowchart on p.~\pageref{flowchart1}).
\item Quintessence DE is completely characterized by its equation of state $w(t)$.  DE physics beyond a canonical scalar field can be probed by searching for an inconsistency between geometry $H(z)$ and growth $\d(k,z)$ (\reffig{tests}).  
\item At the phenomenological level, MG can be probed by searching for fifth forces via comparison of dynamics with lensing or targeting screening effects (\reffig{screening}).  These signatures are very difficult to mimic with DE.
\end{enumerate}
\end{summary}

\section*{ACKNOWLEDGMENTS}
\footnotesize{We thank Alex~Barreira, Dragan~Huterer, Eiichiro~Komatsu, Eric~Linder, and Filippo~Vernizzi for comments, and Paolo~Creminelli for permission to use \reffig{Pk-DE}.  AJ was supported in part by the Kavli Institute for Cosmological Physics at the University of Chicago through grant NSF PHY-1125897, an endowment from the Kavli Foundation and its founder Fred Kavli, and by the Robert R. McCormick Postdoctoral Fellowship.
LL was supported by the STFC Consolidated Grant for Astronomy and Astrophysics at the University of Edinburgh and a SNSF Advanced Postdoc.Mobility Fellowship (No.~161058).
FS acknowledges support from the Marie Curie Career Integration Grant  (FP7-PEOPLE-2013-CIG) ``FundPhysicsAndLSS.''}


%
%
\bibliography{bibfile.bib}
\bibliographystyle{ar-style5.bst}

\end{document}